\documentclass[pra, twocolumn, superscriptaddress, amsmath, amssymb, floatfix]{revtex4-1}

\usepackage[utf8]{inputenc}
\usepackage[english]{babel}
\usepackage{graphicx}
\usepackage{braket}
\usepackage{wrapfig}
\usepackage{hyperref}
\usepackage{bm}
\usepackage{subfigure}
\usepackage{color}
\usepackage{soul}
\usepackage{tikz}
\usepackage{comment}
\usetikzlibrary{arrows}
\def\Tr{\text{Tr}}

  {\left\lbrace\begin{array}{@{}l@{}}}%
  {\end{array}\right.}

\begin{document}

\title{Energy transport between two integrable spin chains}

\author{Alberto Biella}
\email{Corresponding author: alberto.biella@sns.it}
\affiliation{NEST, Scuola Normale Superiore \& Istituto Nanoscienze-CNR, I-56126 Pisa, Italy}
\affiliation{Kavli Institute for Theoretical Physics, University of California, Santa Barbara, CA 93106, USA}

\author{Andrea De Luca}
\affiliation{LPTMS, CNRS, Univ. Paris-Sud, Universit\'e Paris-Saclay, F-91405 Orsay, France}

\author{Jacopo Viti}
\affiliation{Escola de Ci\^encia e Tecnologia da UFRN, Campos Universit\'ario, Lagoa Nova  59078-970 Natal, Brazil}

\author{Davide Rossini}
\affiliation{NEST, Scuola Normale Superiore \& Istituto Nanoscienze-CNR, I-56126 Pisa, Italy}
\affiliation{Kavli Institute for Theoretical Physics, University of California, Santa Barbara, CA 93106, USA}

\author{Leonardo Mazza}
\affiliation{D\'epartement  de  Physique,  Ecole  Normale  Sup\'erieure  /  PSL  Research  University, CNRS,  24  rue  Lhomond,  F-75005  Paris,  France}
\affiliation{Kavli Institute for Theoretical Physics, University of California, Santa Barbara, CA 93106, USA}

\author{Rosario Fazio}
\affiliation{ICTP, Strada Costiera 11, I-34151 Trieste, Italy}
\affiliation{NEST, Scuola Normale Superiore \& Istituto Nanoscienze-CNR, I-56126 Pisa, Italy}
\affiliation{Kavli Institute for Theoretical Physics, University of California, Santa Barbara, CA 93106, USA}

\date{\today}

\begin{abstract}
We study the energy transport in a system of two half-infinite XXZ chains initially kept separated at different temperatures, and later connected and 
let free to evolve unitarily.
By changing independently the parameters of the two halves, 
we highlight, through bosonisation  and time-dependent matrix-product-state simulations, the different contributions of low-lying bosonic modes 
and of fermionic quasi-particles to the energy transport. In the simulations we also observe that the energy current reaches a finite value which 
only slowly decays to zero. The general pictures that emerges is the following.  Since integrability is only locally broken in this model, 
a pre-equilibration behaviour may appear. In particular, when the sound velocities of the bosonic modes of the two halves match, the 
low-temperature energy current is almost stationary and described by a formula with a non-universal prefactor interpreted as a transmission 
coefficient. Thermalisation, characterized by the absence of any energy flow, occurs only on longer time-scales which are not accessible with our numerics.
\end{abstract}

\maketitle

\section{Introduction}

Energy transport in one-dimensional systems is a fundamental problem in non-equilibrium physics. Its understanding, besides 
the clear importance for technological applications~\cite{giazotto2006,li2012}, is tightly bound to a number of  fundamental issues in 
statistical mechanics. The existence of an anomalous transport regime in opposition to the failure of Fourier's law, and its connection 
to the underlying chaotic dynamics were extensively studied in classical interacting systems~\cite{lepri2003,casati2004}. 
In quantum systems, the long standing interest in heat transport in nanowires and one-dimensional edge modes (see 
e.g.~\cite{kane1996,rego1998,fazio1998}) has been recently reinforced by corresponding studies in cold atomic 
systems~\cite{brantut2013}.

\begin{figure}[t]
 \includegraphics[width=\columnwidth]{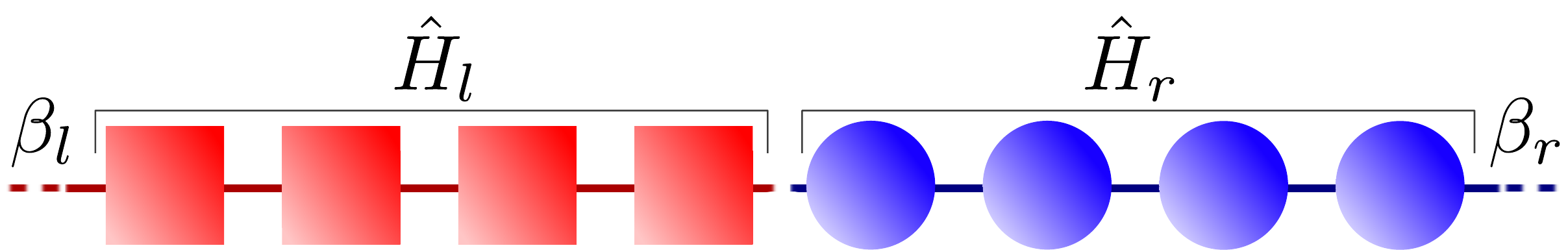}
 \caption{(color online) Sketch of the {\it partitioning} protocol: Two initially
disconnected half chains of semi-infinite length, thermalized
at different inverse temperatures $\beta_l$ and $\beta_r$, are connected at time $t=0$. A net energy current flowing through the junction develops into the system. In this article we investigate the properties of such current when the two halves are integrable (but different) spin chains.}
\label{fig:sketch}
\end{figure}

In the spirit of quantum quenches~\cite{polkovnikov2011}, which are typically implemented in cold-atom setups~\cite{cheneau2012, corman2014, hild2014},  
energy transport in one-dimensional quantum many-body systems can be addressed by means of the following \textit{partitioning} protocol~\cite{Spohn1977, Aschbacher2003, bernard2012, bernard2015}. 
Two (ideally semi-infinite)
chains are prepared in thermal equilibrium with different temperatures (see the sketch in Fig.~\ref{fig:sketch}). The two chains are then connected and the system 
evolves unitarily to a steady state which depends on the initial state and on the total Hamiltonian of the two half-chains and the connecting link. 
A transient regime first appears in which an energy current develops between the two halves.
Then, in the thermodynamic limit, a steady state will be reached and this long-time behaviour may be characterized by a finite or zero energy current.  
Notice that this situation is quite different from the setups usually employed for solid-state devices, where 
a finite-length wire is connected at both ends to two thermal reservoirs.

In the case of homogeneous interacting chains approximated at low-energy by a conformal field theory (CFT), the low-temperature energy current was predicted to be finite and universal in the steady state~\cite{bernard2012, bernard2015}. The result was further confirmed numerically in spin chains~\cite{karr2013} and analytically for free fermions~\cite{deluca2013}. 
While it is clear that this description must apply in some time regimes, in the stationary state the non-universal features of the specific model under consideration might play a major role. In particular, it is not yet clear whether the energy current can be different from zero in the non-equilibrium steady state (NESS) of general interacting spin-chain models.

Earlier works addressing transport in one-dimensional interacting quantum systems within the framework of linear-response theory have pointed out that integrability and in particular the existence of conserved quantities is crucial for the presence of a persistent energy current~\cite{zotos1997}.
There, the persistence of the current was related to an equilibrium Drude weight;
afterwards the relation between integrability and transport properties fully emerged and it was intensively scrutinized in numerous important works (see~\cite{heidrich-meisner2003, jung2006, sirker2011, moore2012, prosen2011-2014} and Refs. therein).
These results call for the understanding of the intimate relation, if any, between the properties of the non-equilibrium steady state (NESS) and the universal features of the evolution, such as integrability, also in the partitioning protocol.

In an early work, it was suggested that the crucial property is not mere integrability  but the presence of a conserved quantity which has non-zero overlap with the energy current~\cite{karr2013}.
Following the ideas presented in Ref.~\cite{CastroAlvaredo2014},
an analytical ansatz for the steady state of the spin-1/2 XXZ model, where the total energy current commutes with the Hamiltonian, was proposed and found in good agreement with numerical simulations~\cite{deluca2014}. 
More recently, under reasonable hypotheses it was demonstrated that when the energy current is a conserved quantity, the energy flow must be ballistic and a non-vanishing stationary current is expected~\cite{Doyon2015}.
For a review of several other peculiar features of the partitioning protocol, see Refs.~\cite{karrash2013, karrash2014, bonnes2014, Viti2015, vasseur2015, DW2015}. 

In its simplicity, the quench protocol considered here poses a number of very interesting questions, some of them still to be answered. 
What are the conditions for the steady-state current to be finite? On the other hand, when is the overall system going to equilibrate to a thermal state?  
Moreover, in one-dimensional spin-1/2 chains bosonization highlights the differences between low-lying bosonic modes and the {high}-energy fermionic quasi-particles.
How do the different scattering properties of these excitations affect the universal low-temperature results for the energy current derived in Ref.~\cite{bernard2012, bernard2015}? How do these results extend to higher temperatures?

In order to contribute to some of these issues, we consider the energy transport between two half-infinite XXZ chains with 
different (and possibly space-dependent) couplings. 
In this setup, we can explore how, at low energies, different interaction parameters and different 
velocities of propagation for the bosonic modes affect the value of the steady-state current. 
Moreover, in this way we are able to study the effect of local integrability-breaking on the dynamics of the energy current. 
In most of the paper we will consider the case in which the couplings are uniform in each half-chain (but with a difference between left and right).
We also consider a smooth crossover between the two halves 
which extends over a finite region. In this way we can also investigate the impact of an adiabatic junction on the scattering of the low-lying modes 
and hence on the energy transport.

What we found is that a pre-equilibration behaviour may set in: the energy current reaches a finite value and then slowly decays toward zero. At low-temperature, this can be justified analytically and is peculiar of the setup we consider. In particular, when the sound velocities of the two halves are matched, the energy current becomes almost stationary and its value is not universal. 
Such pre-thermal regime is still observable at intermediate temperatures by means of numerical simulations.
The asymptotic long-time regime is characterized by the absence of any energy flow. The onset of an effective temperature occurs on longer time-scales not observable with our methods.

The paper is organised as follows. In the next Section we  define in detail the model that we consider, the partitioning protocol anticipated in 
this introduction, and the observables under exam. 
In Section~\ref{sec:low_energy} we discuss the low-temperature regime, where bosonisation leads to an analytic form of the steady-state current. 
Numerical results confirming the previous predictions (but also addressing higher temperatures) are presented in 
Section~\ref{sec:numerics}. These calculations are performed with algorithms exploiting the Matrix Product States (MPS) formalism. 
On the basis of these results, in Section~\ref{sec:scenario} we present a general scenario for energy transport in inhomogeneous quasi-integrable models.
The paper ends with our conclusions.

\section{Transport between two XXZ spin-1/2 chains with different parameters}
\label{sec:problem}

%%%%%%%%%%%%%%%%%%%%%%%%%%%%%%%%%%%%%%%%%%
%
%
%%%%%%%%%%%%%%%%%%%%%%%%%%%%%%%%%%%%%%%%%%%

Let us begin by introducing the XXZ model and the partitioning protocol. 
We consider the Hamiltonian (for an introduction to the properties of this model see Refs.~\cite{Karbach_1998, Lu_2009}):
\begin{align}
 & \hat H = \hat H_l + \hat H_r + \hat h_{l,0},
 \quad
 \hat H_l = \sum_{n<0} \hat h_{l,n},
 \quad 
 \hat H_r = \sum_{n>0} \hat h_{r,n}, \nonumber
 \\
 & \hat h_{\lambda,n} = J_\lambda [\hat S_{n}^x \hat S_{n+1}^x
 + \hat S_{n}^y \hat S_{n+1}^y
 + \Delta_\lambda \hat S_{n}^z \hat S_{n+1}^z
 ] ,
 \label{eq:Hamiltonian}
\end{align}
$\hat S^\alpha_n$ being the $\alpha$-th component of the spin-$1/2$ 
on site $n$ ($\hbar=k_{\rm B}=1$) and $\lambda =l,r$.
We restrict ourselves to the critical phase $-1<\Delta_\lambda \leq 1$ (as this is most relevant for the energy transport) and 
assume $J_\lambda > 0$. At the beginning the system is separated into two independent halves
held at different inverse temperatures $\beta_{l,r}=1/T_{l,r}$
\begin{equation}
\label{initial_state}
 \hat \rho_0 = \mathcal Z^{-1} e^{-\beta_l \hat H_l} \otimes 
 e^{-\beta_r \hat H_r},
\end{equation}
where $\mathcal Z$ ensures the normalization of the density matrix.
For times $t \geq 0$ the state is unitarily evolved with the Hamiltonian
$\hat H$ defined in Eq.~\eqref{eq:Hamiltonian}, so that the initially-separated reservoirs are 
put in contact. A heat flow is generated, due to the temperature unbalance in the
initial conditions. 

The aim of this article is the investigation of the energy 
flow close to the junction (site $n=0$) in the long-time regime:
\begin{align}
 \mathcal J =& \lim_{t \to \infty} \mathcal J(t) =
 \lim_{t \to \infty} \text{Tr} [\hat \rho(t) \, \hat j_0],
 \\
 \hat j_0 =& \frac i2 [\hat H, \hat H_r-\hat H_l].
 \label{eq:current0}
\end{align}
In order to point out the existence of a stationary energy current in one-dimensional systems, it is customary to employ arguments based on the linear-response theory. 
These arguments are not applicable here, so that we do not expect any ballistic transport of energy at large times, $\mathcal J=0$.
In particular, we provide evidence that the system is non-integrable and that no conserved quantity prevents the current from decaying to zero.

We first observe that $\hat H_l$ and $\hat H_r$ describe two XXZ models with
different parameters $J_{l,r}$ and $\Delta_{l,r}$. They are integrable, as 
confirmed by the Poissonian level-spacing statistics and the existence 
of an infinite number of conserved local operators. In particular, the energy 
current operator
\begin{equation}
 \hat j_{\lambda,n} = i  [\hat h_{\lambda,n-1}, \hat h_{\lambda,n}]
 \label{eq:energy:current:density}
\end{equation}
is a conserved density because of the following continuity equation:
\begin{equation}
 \partial_t \hat j_{\lambda,n} = i[\hat H_\lambda, \hat j_{\lambda,n}] = \hat k_{\lambda,n-1}-\hat 
k_{\lambda,n}.
\end{equation}
Thus, the integrated energy current from site $a$ to site $b$ $\sum_{n=a}^b \hat j_{\lambda,n}$ 
changes in time only because of the boundary terms $\hat k_{\lambda,a}$ and
$\hat k_{\lambda,b}$. Operators $\hat k_{\lambda,n}$ are usually interpreted as \textit{longitudinal pressures}~\cite{prosen2011-2014}.
Additionally, the two Hamiltonians are exactly solvable with Bethe ansatz even 
in presence of a hard-wall boundary, as in this case.
Even if the Hamiltonian $\hat H$ in Eq.~\eqref{eq:Hamiltonian} is obtained by 
joining two integrable models, whenever $\Delta_l \neq \Delta_r$ or $J_l 
\neq J_r$ (with $\Delta_{l} \neq 0$ or $\Delta_{r} \neq 0$) the integrability is broken. This can be directly observed by inspection 
of the level-spacing statistics, which already at finite size suggests an 
abrupt change to a Wigner-Dyson distribution (see Appendix~\ref{app:levstat} 
for more details). 
{We note that defining integrability is a subtle problem in quantum mechanics~\cite{Caux:Int}. Here, we refer to the notion of integrability associated to Bethe ansatz solvability, for which the level spacing statistics has been shown to be Poissonian~\cite{polkovnikov2011, Canovi:2011}.}
At the same time, this model is not known to display any 
non-trivial local conserved quantity beyond the total energy and magnetization. 
Thus, there seems to be no element in the dynamics that supports a stationary current~\cite{zotos1997, polkovnikov2011, karrash2013, Doyon_arxiv2015}.

However, the original properties of the two halves translate into just a \textit{local} source of integrability breaking. Let us consider for example the energy-current operator: 
\begin{equation}
 \hat j_{a,b} = \sum_{n=a}^{-1} \hat j_{l,n} +
 2\hat j_0 +
 \sum_{n=2}^b \hat j_{r,n},
 \qquad a<0, \; b >0.
 \label{eq:jabdef}
\end{equation}
The cases where $a>0$ or $b<0$ can be easily deduced but are not considered 
explicitly here. 
The time derivative of $\hat j_{a,b}$ reads:
\begin{equation}
 i [\hat H, \hat j_{a,b}] = \hat k_{l,a} - \hat k_{r,b} + \hat \Theta
 \label{eq:Theta}
\end{equation}
where $\hat \Theta$ is a local operator with support on a few sites around 
$n=0$ (see Appendix~\ref{app:theta} for the explicit formula).

When $\hat \Theta =0$, the fact that $\hat k_{l,a}$ and $\hat k_{r,b}$ are localized on a few sites which are far from the junction allows for the derivation of a lower bound on $\mathcal J$ which is valid beyond linear response theory. This has been explicitly carried out {in a general situation which includes the case} when $\Delta_l=\Delta_r$ and $J_l = J_r$~\cite{Doyon2015}.
The argument links the local energy current $\mathcal J$ to $\hat j_{a,b}$ for $b,- a \gg 1$ assuming a 
sufficiently regular transient behaviour of $\hat j_{\lambda,n}$ along the chain.
Even if this argument cannot be applied to our case of interest, the locality 
of the $\hat \Theta$ operator may allow the persistence of a quasi-stationary current in the long-time regime.
In the following we will see that a pre-equilibration behavior emerges with non-zero energy current, which in some regimes can be quantitatively predicted. 
Thermalisation and in particular the absence of energy flows ($\mathcal J = 0$) occur only at later stages which are not numerically accessible. We will characterize them by general arguments.
{
This does not happen in the special case $\Delta_l = \Delta_r$ and $J_l = J_r$, which has been the subject of extensive numerical and analytical studies~\cite{karrash2014, deluca2014}.
In this case there is no occurrence of thermalisation even at very large times, and, in particular, the stationary state depends strongly on the initial condition in the two halves.
}

\section{Low-energy limit}
\label{sec:low_energy}

We begin by presenting an analytical formula for $\mathcal J(t)$ in the pre-equilibration limit. 
In the low-temperature regime, one can exploit the low-energy (LE) field-theory and bosonisation to derive a transmission 
coefficient which originates from the inhomogeneity of the Hamiltonian $\hat H$ in Eq.~\eqref{eq:Hamiltonian} and generalises the formula obtained for the
homogeneous case in Refs.~\cite{bernard2012, bernard2015}.

We start rewriting the Hamiltonian density
$\hat{h}_{\lambda, n}$ in Eq.~\eqref{eq:Hamiltonian} 
in terms of canonical local fermionic operators by standard Jordan-Wigner transformation as
\begin{multline}
\label{eq:fermionicJWXXZ}
 \hat{h}_{\lambda,n}=-\frac{J_{\lambda}}{2}\bigl(\hat c^{\dagger}_{n+1} \hat c_n 
 + \mathrm{h. c.}) +\\
 +J_{\lambda}\Delta_{\lambda}
  \hspace{-0.1cm} \left(\hat c^{\dagger}_n \hat c_n -\frac{1}{2} \right)\hspace{-0.1cm} \,
 \hspace{-0.1cm} \left(\hat c^{\dagger}_{n+1}\hat c_{n+1}-\frac{1}{2}\right) \hspace{-0.1cm} ~.
\end{multline}
Since the inhomogeneous model preserves the
total magnetization and for $|\Delta_{\lambda}|<1$ its ground state $|GS\rangle$ is still in the zero 
magnetization  sector {($\langle GS|\sum_n \hat{S}_n^z|GS\rangle=0$)},  we have $\langle GS|\hat c^{\dagger}_n \hat c_n|GS\rangle=1/2$.  Thus we can rewrite the Hamiltonian \eqref{eq:fermionicJWXXZ} as
\begin{equation}
\label{eq:fermionicXXZ}
 \hat{h}_{\lambda,n}=-\frac{J_{\lambda}}{2}\bigl(\hat c^{\dagger}_{n+1} \hat c_n 
 + \mathrm{h. c.}) +
 J_{\lambda}\Delta_{\lambda}
 : \hspace{-0.1cm} \hat c^{\dagger}_n \hat c_n \hspace{-0.1cm} : \,
 : \hspace{-0.1cm} \hat c^{\dagger}_{n+1}\hat c_{n+1} \hspace{-0.1cm} :~.
\end{equation}
where the notation $: \hspace{-0.1cm} \hat c^{\dagger}_n \hat c_n \hspace{-0.1cm} :$ above stands for the normal ordering prescription
$: \hspace{-0.1cm} \hat c^{\dagger}_n \hat c_n \hspace{-0.1cm} : = \hat c^{\dagger}_n \hat c_n-\langle GS| \hat c^{\dagger}_n \hat c_n|GS\rangle$.
In order to get a low-energy theory we expand 
the local fermionic operators in terms of continuous
fields. Starting from the non-interacting theory,
we describe the excitations 
near the Fermi momentum $k_F=\pi/2a$, where $a$ is the lattice spacing, $x=na$. 
Then, it is possible to expand the operator $\hat c_n$ in the following way
\begin{equation}
 \frac{\hat c_n}{\sqrt{a}}\simeq e^{ik_F x}\hat\psi_+(x)+e^{-ik_Fx}\hat\psi_-(x),
\end{equation}
where $\hat\psi_{\pm}$ are the chiral Dirac fields. 
This procedure is described in
details in Ref.~\cite{aff2014} for general
non-translational invariant Hamiltonians of the type \eqref{eq:Hamiltonian} allowing for site-dependent
parameters.

Subsequently, the fermionic fields are
bosonised as $\hat\psi_{\pm}(x)=e^{\pm i\sqrt{4\pi}\hat\phi_{\pm}(x)}$, with bosonic action
\mbox{$S=1/2\int~dt dx(\partial_\mu \phi)^2$}, where $\phi = \phi_+ + \phi_-$. 
The effective low-energy theory is in our case specified by the Hamiltonian
$\hat H_{LE}=\hat H_{LL}+\hat V$. The first piece $\hat H_{LL}$ is the inhomogeneous Luttinger liquid (LL) Hamiltonian
\begin{equation}
\label{eq:LL_in}
\hat H_{LL}=\frac{1}{2}\int_{-\infty}^{\infty}dx u(x)~\left[\frac{(\partial_x \hat \phi)^2}{K(x)}+K(x)(\partial_x\hat\theta)^2\right],
\end{equation}
with fields obeying $[\hat \phi(x,t),\partial_x\hat\theta(x',t)]=i\delta(x-x')$ and $K(x)\partial_x\hat\theta=\partial_t\hat\phi$. 
The parameters $K(x)$ and $u(x)$ are the Luttinger parameter and the renormalised Fermi velocity. For an abrupt junction 
$K(x)=K_l$ for $x<0$ and $K(x)=K_r$ for $x\geq 0$, analogously $u(x)=u_l$ for negative values of $x$ and $u(x)=u_r$ for $x\geq 0$. Those parameters are related to the
lattice coupling constants in \eqref{eq:Hamiltonian} as follows 
\begin{equation}
 K_{\lambda}=\frac{1}{2}\Bigl[ 1-\frac{1}{\pi}\arccos\Delta_{\lambda}\Bigr]^{-1},
\end{equation}
and $u_\lambda=\nu_{\lambda}K_{\lambda}/(2K_{\lambda}-1)$ with $\nu_{\lambda}=J_{\lambda}\sqrt{1-\Delta_{\lambda}^2}$.

%%%%%%%%%%%%%%%%%%%%%%%%%%%%%%%%%%%%%%%%%%%
%
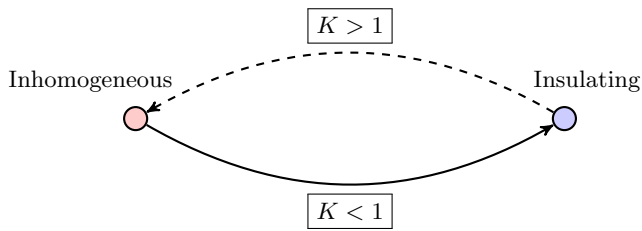
\begin{figure}
\centering
\begin{tikzpicture}[->,>=stealth',auto,node distance=5.7cm,
  thick,main node/.style={circle,draw,font=\sffamily\Large\bfseries}]

  \node[main node, fill=red!20] (1) {};
  \node[main node, fill=blue!20] (2) [right of=1] {};

  \path[every node/.style={font=\sffamily\small}, thick]
    (1) edge[bend right] node [below] {$\boxed{K<1}$} (2);
    \path[every node/.style={font=\sffamily\small}, thick, dashed]
    (2) edge[bend right] node [above] {$\boxed{K>1}$} (1);
\node at (-0.6,0.5) {Inhomogeneous};
\node at (6,0.5) {Insulating};
 
\end{tikzpicture}
\caption{(color online) The renormalization group flow in presence of the backscattering operator \eqref{eq:perturbation}. The
inhomogeneous fixed point  is described by a CFT  and is conducting, with energy
current given in \eqref{eq:current_CFT}. At the insulating  fixed point  the renormalized backscattering coupling constant $\lambda$ is formally infinite and one should expect absence of thermal
transport at low energy.}
\label{fig:RGflow}
\end{figure}
%
%%%%%%%%%%%%%%%%%%%%%%%%%%%%%%%%%%%%%%%%%%%

The perturbation $\hat V$ is the localised backscattering operator,
responsible for reflection of fermionic waves through the junction at
$x=0$, and is given by~\cite{aff2012, aff2014}
\begin{equation}
\label{eq:perturbation}
 \hat V=\lambda (e^{i\sqrt{4\pi}\hat \phi(x=0)}+\rm{h.c.}).
\end{equation}
For an abrupt junction~\cite{aff2012, aff2014}, the  coupling $\lambda$ is proportional to the difference between the renormalized Fermi 
velocities on the two sides of the chain,
i.e. $\lambda\sim (u_l-u_r)$. Its relevance at the fixed point described by the inhomogeneous LL Hamiltonian \eqref{eq:LL_in}  depends on the parameter
\begin{equation}
 K=2\left(\frac{1}{K_l}+\frac{1}{K_r}\right)^{-1},
 \label{eq:Kdef}
 \end{equation}
in close analogy with the seminal work \cite{kane1996}.
It is relevant when $K<1$, marginal at $K=1$, and irrelevant for  $K>1$.  
Equating the two renormalised Fermi velocities $u_l=u_r$ we can directly set the coupling $\lambda=0$ 
and explore the inhomogeneous LL fixed point.  The other fixed point of the renormalization group (RG) flow is characterised by a 
formally infinite backscattering of the fermionic
modes and should be insulating, see Fig.~\ref{fig:RGflow}. The numerical analysis of Sec. \ref{sec:numerics} shows in this 
case a slow decay of the energy current.

The complete study of the flow between these
two fixed points at finite temperatures with techniques analogous to those in~\cite{sal} is problematic;
in the following we limit ourselves to the derivation of the explicit form of the energy current
at the inhomogeneous LL fixed point,
relying on the techniques developed in~\cite{Viti2015}. {We will
also not make any attempt to compute higher cumulants of the energy current as discussed in a similar
field theoretical framework for  charge transport in~\cite{sal_charge}.}

{Determining the current requires} to study  the long-time dynamics of two different conformal field theories (free bosons with different 
compactification radii) at different temperatures {when}
the interface, or defect, preserves conformal symmetry. 
The Appendix~\ref{app:CFT} contains the technical details of the calculation for the interested reader,
the result for the energy current is
\begin{equation}
\label{eq:current_CFT}
 \mathcal{J}_{LE}
 =
 \frac{\pi \mathcal T}{12}(\beta_l^{-2}-\beta_r^{-2}),
\end{equation}
where the transmission coefficient is
\begin{equation}
\label{eq:transmission}
 \mathcal T=\frac{4\alpha}{(1+\alpha)^2},\quad\alpha=\frac{K_l}{K_r}.
\end{equation}

%%%%%%%%%%%%%%%%%%%%%%%%%%%%%%%%%%%%%%%%%%%
%
\begin{figure}[t]
 \includegraphics[width=\columnwidth]{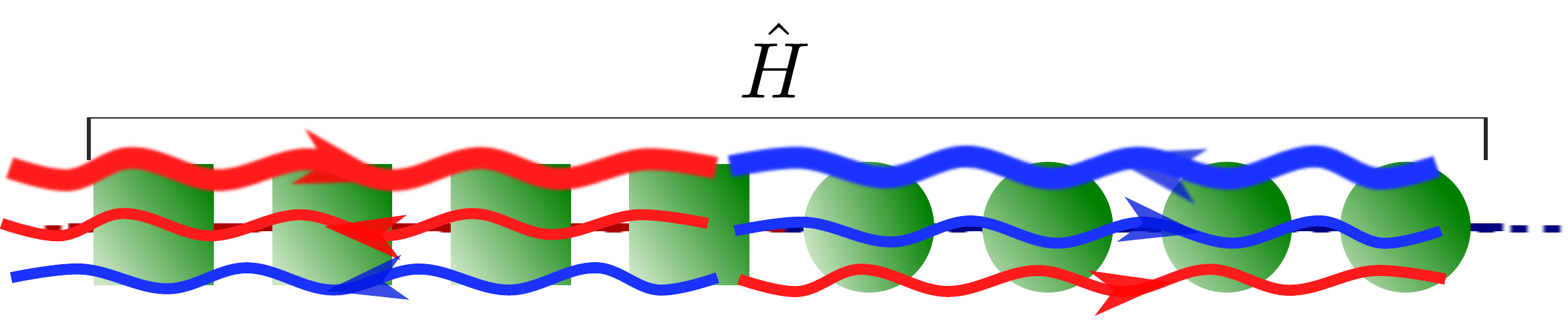}
 \caption{
 (color online) Sketch of the stationary state according to the low-energy Luttinger-liquid description: 
 when the Fermi velocity on the two sides are matched, the discontinuity of the Luttinger parameter
 $K(x)$ induces transmission/reflection coefficients: the hot (cold) low-lying bosonic modes coming
 from left (right) are reflected and trasmitted at the junction, determining a non-universal value for the 
 stationary energy current.
\label{fig:sketch:2}}
\end{figure}
%
%%%%%%%%%%%%%%%%%%%%%%%%%%%%%%%%%%%%%%%%%%%

The coefficient $\mathcal T$ has a simple physical interpretation: It is the transmission coefficient associated to the scattering of the 
bosonic low lying modes past the step in the coupling constants at the site connecting the two half-chains.
It can be  obtained by
solving the classical equation of motion for the bosonic field $\phi(x,t)$ (see the sketch in Fig.~\ref{fig:sketch:2}). Such an equation can be
obtained from the Lagrangian that follows from \eqref{eq:LL_in} having set $u_l=u_r$ and reads
\begin{equation}
\label{eq:eom}
 \partial_t^2\phi(x,t)=K(x)\partial_x\left[\frac{\partial_x\phi(x,t)}{K(x)}\right]
\end{equation}
with boundary conditions $\phi(0^+,t)=\phi(0^-,t)$ and 
\begin{equation}
\label{eq:LR_t}
 \frac{\partial_x \phi(0^-,t)}{K_l}=\frac{\partial_x \phi(0^+,t)}{K_r}.
\end{equation}
The above condition ensures continuity of the momentum density $
p(x,t)= \frac{1}{K(x)}\partial_t\phi\partial_x\phi$ at $x=0$.
We can look for a plane-wave solution of the equation of motion~\eqref{eq:eom} and~\eqref{eq:LR_t} of the form $\phi(x,t)=e^{ik(x-t)}+\sqrt \mathcal Re^{-ik(x+t)}$ for
$x<0$ and $\phi(x,t)=  \sqrt \mathcal T e^{ik(x-t)}$ for $x>0$ and solve for the reflection and transmission coefficients $\mathcal R$ and $\mathcal T$ with $\mathcal R+\mathcal T =1$. The transmission coefficient
which is obtained coincides with~\eqref{eq:transmission}.

%%%%%%%%%%%%%%%%%%%%%%%%%%%%%%%%%%%%%%%%%%%%%%%%%%%%%%%%%%%%%%%%% 
%
\begin{figure}[t]
\centering
\includegraphics[width=0.4\textwidth]{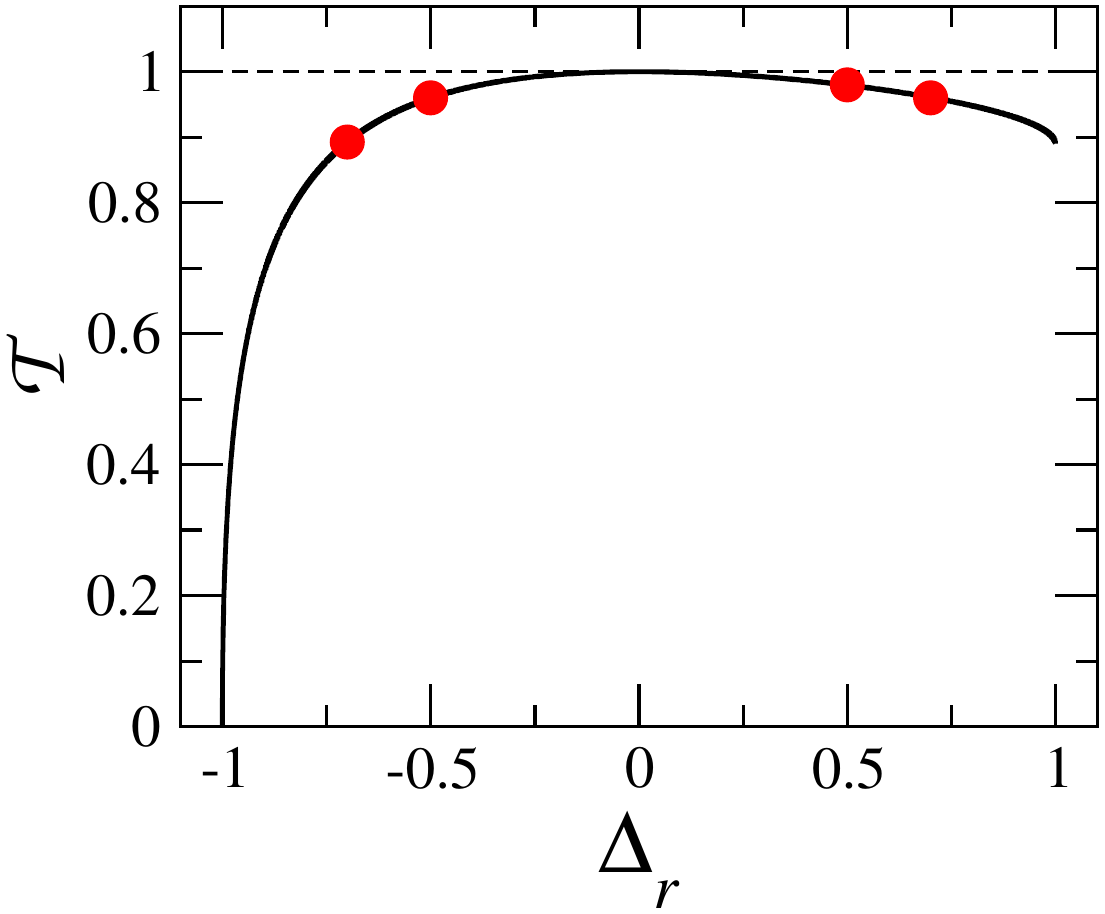}
\caption{(color online) Transmission coefficient $\mathcal{T}$ as a function of the anisotropy parameter $\Delta_r$ for $\Delta_l=0$. 
The red dots signal the parameters used in the simulations displayed in Fig.~\ref{fig:XXZ_scalingT}.
The dashed line indicates the upper limit $\mathcal{T}=1$.}
  \label{fig:t_dl0}
\end{figure}
%
%%%%%%%%%%%%%%%%%%%%%%%%%%%%%%%%%%%%%%%%%%%%%%%%%%%%%%%%%%%%%%%%% 

Even if the expression for the current  is corrected by a transmission coefficient,  it still retains the simple functional form
\begin{equation}
\label{num01}
\mathcal J_{LE} = f(\beta_l) - f(\beta_r),
\end{equation}
firstly observed numerically in Ref.~\cite{karr2013}. The CFT analysis assumes a spectrum with 
linear dispersion relation where all the particles travel at the same speed. The curvature of the lattice dispersion, or from a field theory
perspective irrelevant perturbations, may alter especially at higher temperatures the intuitive picture of a homogeneous NESS  inside the light-cone 
spreading out ballistically. The major consequence will be the emergence of a truly inhomogeneous current profile~\cite{vasseur2015, DW2015} possibly describable by a CFT in a curved space-time~\cite{arctic2015}.  Conscientiously the prediction (\ref{eq:current_CFT})  holds at fixed spacial coordinate $x$ after having waited a time $t$
such that $x/t\rightarrow 0$.

The mechanism for thermalisation at larger times cannot  be captured by the low-energy theory that approximates the original non-integrable Hamiltonian
(Eq.~\eqref{eq:Hamiltonian}) with a free bosonic field theory. 
Nevertheless, we expect~\eqref{eq:current_CFT} to accurately describe the pre-equilibration behavior
near the junction which, as we will see, emerges at times $t \sim J_l^{-1}$.  The robustness of ballistic energy transport is numerically confirmed at short times,
see Sec. \ref{sec:numerics}, also in the presence of a relevant backscattering perturbation. 
{Finally notice that for a smooth junction, the anisotropy parameter is constant
over a length scale of one single lattice site. Bosonization can follow the same procedure as in the homogenenous case where the backscattering operator \eqref{eq:perturbation}
is ruled-out precisely by such a symmetry, see for example~\cite{Affleck_89}.}

\section{Numerical analysis}
\label{sec:numerics}

%%%%%%%%%%%%%%%%%%%%%%%%%%%%%%%%%%%%%%%%%%%%%%%%%%%%%%%%%%%%%%%%% 
%
\begin{figure*}[t]
\centering
\includegraphics[width=\textwidth]{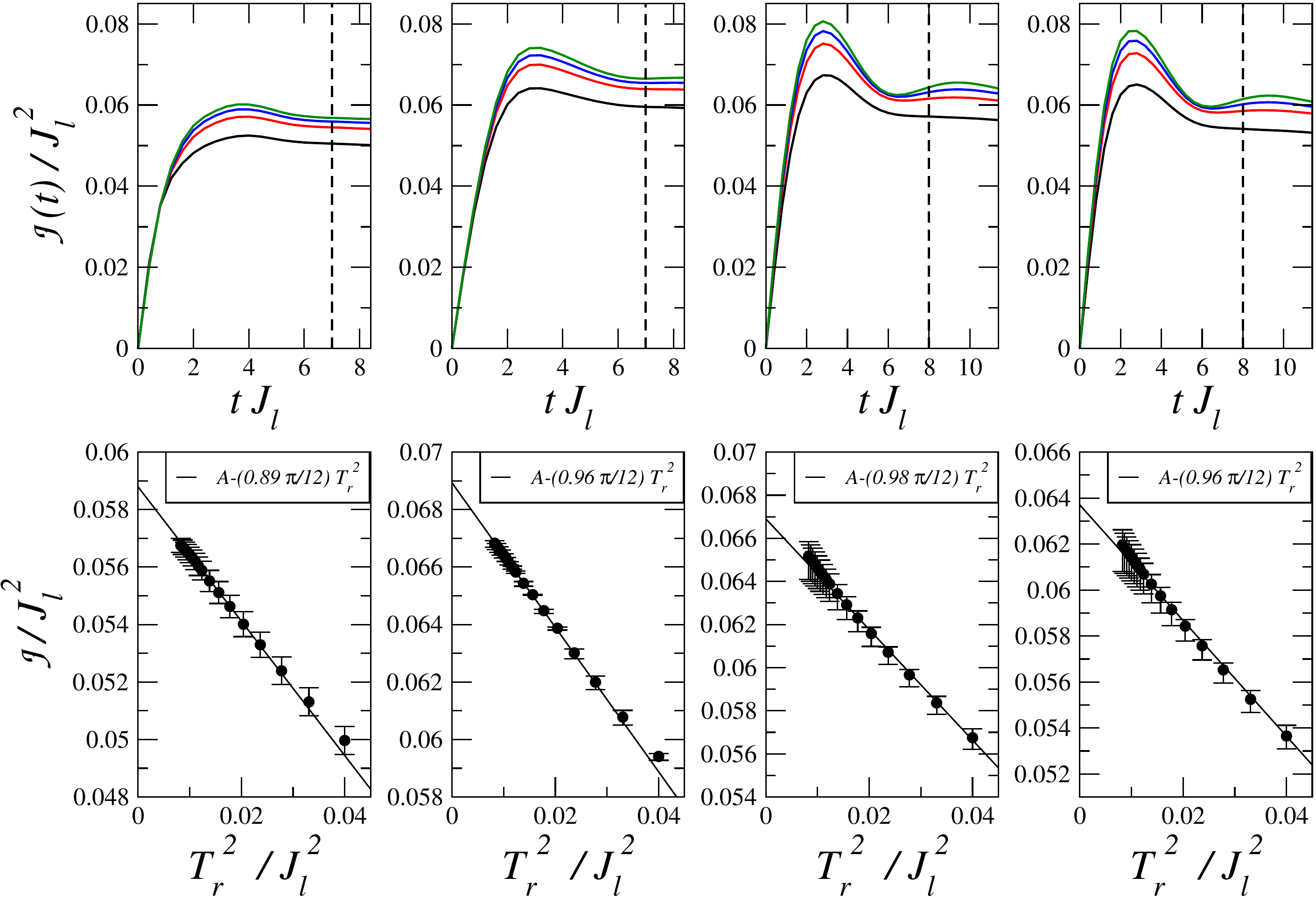}
\caption{(color online) Top panels: the time evolution of energy current $\mathcal J(t)$ for different values of $\beta_r$, from bottom to top: $5, 7, 8.5, 10.5 \ J_l^{-1}$ at fixed $\beta_l=0.1J_l^{-1}$.
Bottom panels: steady-state value of $\mathcal J(t)$ obtained averaging over times \mbox{$t \ge t_{min}$} (dashed vertical line in the top panels).
The parameters of the right halves are (from left to right) \mbox{$J_r/J_l=2.092, 1.54, 0.77, 0.709$} and \mbox{$\Delta_r=-0.7,-0.5,0.5,0.7$} corresponding to \mbox{$\mathcal{T}=0.89,0.96,0.98,0.96$}. In all the cases $\Delta_l=0$ and $u_l=u_r= J_l$.}
  \label{fig:XXZ_scalingT}
\end{figure*}
%
%%%%%%%%%%%%%%%%%%%%%%%%%%%%%%%%%%%%%%%%%%%%%%%%%%%%%%%%%%%%%%%%% 

In order to complement  the results detailed in the previous section and extend them outside the low-temperature limit, we analyze
the problem numerically.
The algorithms we use are detailed in Appendix~\ref{ssec:method} and employ a MPS representation of the purified thermal and time-evolved state~\cite{white2005, schollwock}. By means of a time-evolving block-decimation (TEBD) procedure~\cite{Vidal2004, Daley_2004, White_2004}, we are able to compute the time-evolution of the thermal system~\cite{moore2012, kennes2014} and thus of the current flowing at the junction.

In Sec.~\ref{ssec:match} we test the predictions of the low-energy effective theory developed in the previous Section, and in particular the validity of the formula~\eqref{eq:current_CFT} when $u_l=u_r$.
The more general case $u_l\neq u_r$ is considered in Sec.~\ref{ssec:nomatch}, where we also discuss differences and analogies with the previous case.

\subsection{Fermi-velocity matching case ($u_l=u_r$)}
\label{ssec:match}

In this subsection we want to validate the predictions of the inhomogeneous Luttinger liquid under the Fermi velocity matching condition $u_l=u_r$ in Eq.~\eqref{eq:transmission}.
The fact that the factorization of the energy current approximately holds also in the inhomogeneous case has been numerically verified in Ref.~\cite{karr2013} in a wide range of temperatures. This allows us to verify Eq.~\eqref{eq:transmission} preparing just one of the two halves at low temperature (in our case the right one) irrespectively of the temperature of the other half, a trick which allows us to explore longer times. 

We simulate a free-fermion chain ($\Delta_l=0$) coupled to an interacting one ($\Delta_r=-0.7, -0.5, 0.5, 0.7$).
The hopping parameter of the right half $J_r$ is chosen in order to fulfill $u_l=u_r=J_l$.
The left chain is initially prepared at high temperature $\beta_l=0.1 J_l^{-1}$ while the right chain is initialized at low temperature 
$\beta_r J_l \in [ 5 ,11]$.
In Fig.~\ref{fig:t_dl0} we plot the theoretical prediction of the transmission coefficient $\mathcal{T}$.
Unfortunately, the study of systems where the transmission coefficient is significantly different from $1$, and thus the presence of a transmission coefficient is most clearly verified, proved to be beyond our numerical possibilities. In particular, in the cases where $\Delta_r \sim -1$ (see Fig.~\ref{fig:t_dl0}), the time-evolution is significantly slower {(this is due to the fact that an increasing mismatch of the microscopic parameters of the two halves makes the state more complex at  the junction)} and does not allow a reliable assessment of the quasi-stationary behaviour because of a longer transient dynamics.

In Fig.~\ref{fig:XXZ_scalingT} we present our numerical results. In {all} the cases we observe that after a transient time  the energy current displays a quasi-stationary value, reminiscent of the physics of integrable homogeneous models. 
However it should be stressed that, for the accessible times, the energy current has a residual time dependence {(more evident in the $\Delta_r>0$ cases)} which makes the extrapolation of its quasi-stationary value less accurate.
For this reason, the quasi-stationary value of the current is obtained after averaging over times $t \ge t_{min}$ {(dashed vertical line in the top panels of Fig.~\ref{fig:XXZ_scalingT})}. The value is then plotted as a function of $T_r^2$. 
{The error bars quantify the deviations from the average value, they are evaluated considering the maximum and minimum values of the energy current for $t \ge t_{min}$.}
Our numerical data confirm that at low temperatures  the energy current scales with $T_r^2$ and the slope is compatible with the theoretical prediction of the transmission coefficient in Eq.~\eqref{eq:transmission}. 

The times that we are able to access for this setup are limited and it is very hard to make definite statements about the stationary values of the 
current. While the time increases,
the correlations spread inside a light cone determined by the Fermi velocities.
From the numerical point of view, this implies that the bond link $\chi$ needed to faithfully represent the state of the system increases exponentially in time (see Appendix~\ref{ssec:method}).
For this reason, our numerics is not effective in exploring the time-scale of thermalisation for which the energy current vanishes.
  
\subsection{Generic case ($u_l\neq u_r$)}
\label{ssec:nomatch}

In this subsection we consider the generic case where the parameters of the two chains are chosen such that $u_l\neq u_r$.
As discussed in Sec.~\ref{sec:low_energy}, the low-energy analysis tells us that in this case the backscattering operator performs a non-trivial renormalization flow, pointing out the prominent importance of the parameter $K$ in Eq.~\eqref{eq:Kdef}.
Our simulations show that at low temperatures the behaviour of the energy current is qualitatively different from the $u_l=u_r$ case.

%%%%%%%%%%%%%%%%%%%%%%%%%%%%%%%%%%%%%%%%%%%
%
\begin{figure}[t]
\centering
\includegraphics[width=0.40\textwidth]{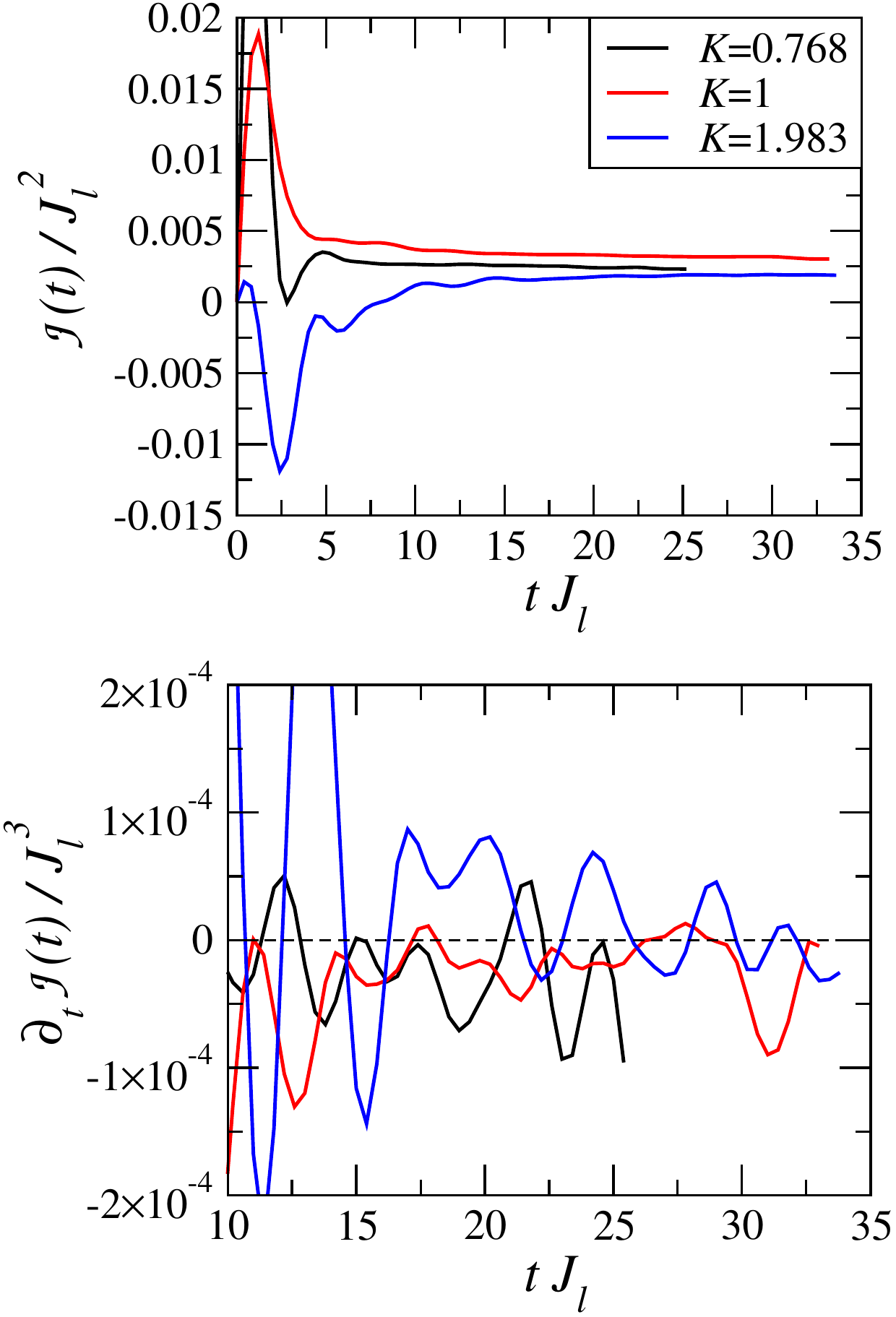}
\caption{Time evolution of the energy current $\mathcal J(t)$ (top panel) and of its derivative (lower panel) for $K=0.768$ (black line), $1$ (red line), $1.983$ (blue line) in the $u_l\neq u_r$ case.
We set $\beta_l = 4 J_l^{-1}, \beta_r = 5 J_l^{-1}$, $\Delta_l=-0.3$, $J_r/J_l=1$ and $\Delta_r=0.95, 0.3, -0.95$ (black, red and blue line respectively).}
\label{fig:derivativeG}
\end{figure}
%
%%%%%%%%%%%%%%%%%%%%%%%%%%%%%%%%%%%%%%%%%%%%%%%%%%%%%%%%%%%%%%%%% 

%%%%%%%%%%%%%%%%%%%%%%%%%%%%%%%%%%%%%%%%%%s%%%%%%%%%%%%%%%%%%%%%%% 
\begin{figure}[t]
\centering
\includegraphics[width=0.40\textwidth]{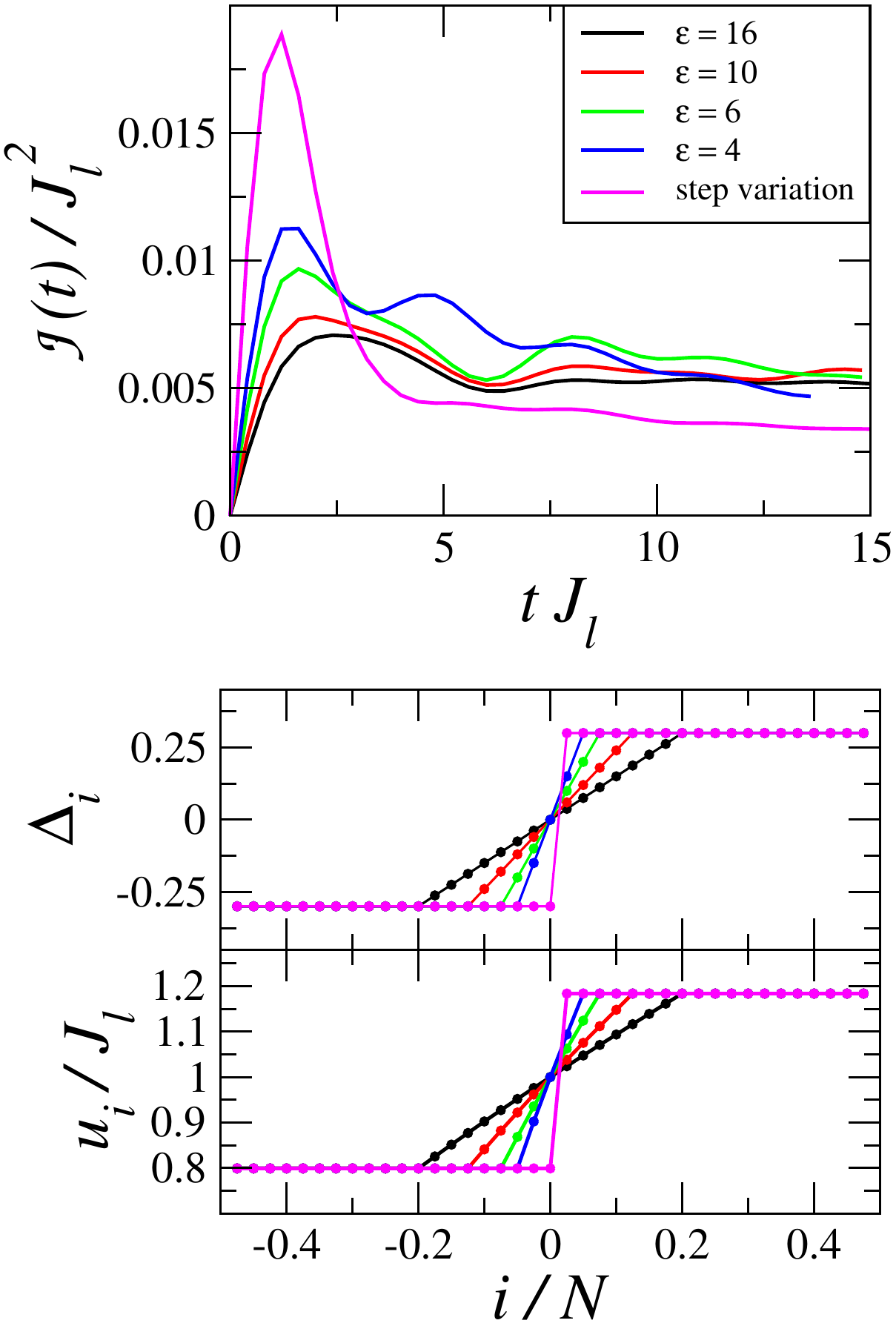}
\caption{(color online) Top panel: time evolution of the integrated energy current $\mathcal J(t)$ for different values of $\varepsilon$ as indicated in the legend.
Bottom panels: local anisotropy parameter $\Delta_i$ and local Fermi velocity $u_i$ as a function of the position.
Here the parameters are $\beta_l=4 \ J_l^{-1}$, $\beta_r=5 \ J_l^{-1}$, $\Delta_l=-\Delta_r=-0.3$ and $J_r/J_l=1$.
}
  \label{fig:ramp}
\end{figure}
%%%%%%%%%%%%%%%%%%%%%%%%%%%%%%%%%%%%%%%%%%%%%%%%%%%%%%%%%%%%%%%%% 

In Fig.~\ref{fig:derivativeG} we present our results for three cases characterized by $K<1$, $K=1$ and $K>1$. The parameters are chosen so that the Fermi velocities of the simulations are of order $J_l$, so that results are comparable for the considered time scales.
Even if this numerical analysis is not conclusive (because the longest accessible times are not enough to obtain a full picture of the pre-thermalisation and thermalisation dynamics), it is still possible to make some interesting observations.
Looking at $\mathcal J(t)$ in Fig.~\ref{fig:derivativeG}~(top) and at its time derivative in Fig.~\ref{fig:derivativeG}~(bottom), we can identify two different behaviours. The former characterizes the cases $K<1$ and $K=1$ and displays steady and slow decay. This result is consistent with the prediction that the system should flow to an insulating fixed point where no energy can be transmitted. 
The latter appears for $K>1$ and displays a current which increases with time. Again, this is consistent with the renormalization-group prediction which states that the perturbation at the center is irrelevant.

Let us finally consider the fact that the perturbation $\hat{V}$ in Eq.~\eqref{eq:perturbation} is the result of the abruptness of the junction and would not appear if $\Delta$ and $J$ would vary smoothly in space (see Sec.~\ref{sec:low_energy}).
To this aim, we numerically investigate a system where the anisotropy parameter is space dependent, $\Delta_i$, and linearly interpolates between the values $\Delta_l$ and $\Delta_r$ in a region of length $\varepsilon$ located in the center of the system. 
$\mathcal J(t)$ is shown for several values of $\varepsilon$ in Fig.~\ref{fig:ramp}~(top).
As $\varepsilon$ increases, and thus the junction is smoothened, the time dependence of the energy current transforms from an apparent decay behaviour to a more quasi-steady one.
Intuitively, the smooth variation of $\Delta_i$ in space implies a smooth variation of the local Fermi velocity $u_i$, as shown in Fig.~\ref{fig:ramp}~(bottom). Since we have pointed out that matching the Fermi velocities is a sufficient condition to guarantee the irrelevance of the perturbation at the junction, this is consistent with the observation of a non-decaying current $\mathcal J(t)$ for large values of $\varepsilon$.

\section{General scenario}
\label{sec:scenario}

%%%%%%%%%%%%%%%%%%%%%%%%%%%%%%%%%%%%%%%%%%%%%%%%%%%%%%%%%%%%%%%%% 
%
\begin{figure}
\includegraphics[width=0.40\textwidth]{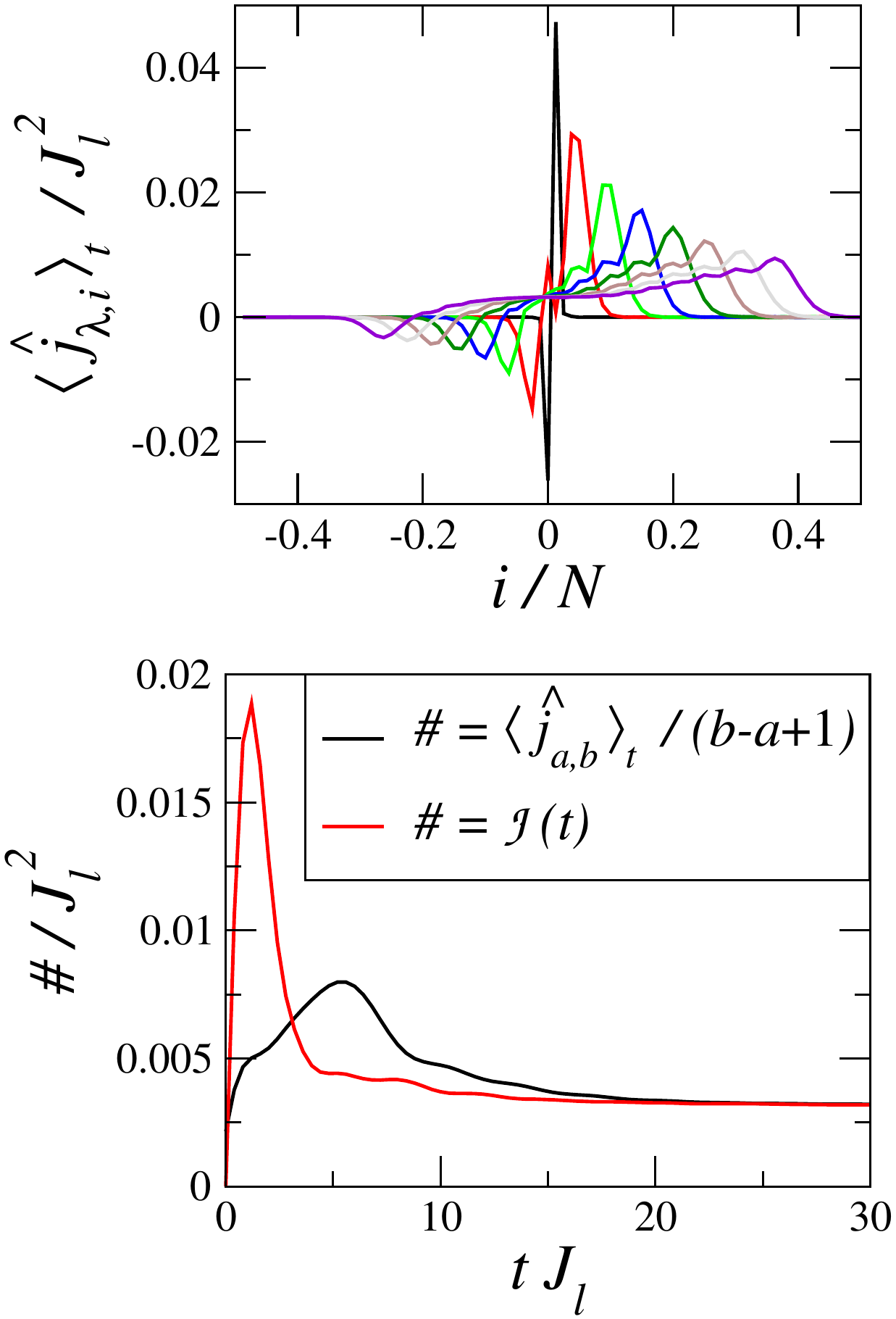}
\caption{(color online) Top panel: snapshots of the local energy current $\braket{\hat j_{\lambda,i}}_t$ at $t J_l / 0.4 = 1, 11, 21, 31, 41,  51, 61$.
Lower panel: comparison between the energy current flowing at the center of the chain and
the integrated energy current for $a=-5$ and $b=+5$ normalized by ($b-a+1$).  
We set $\beta_l = 4 J_l^{-1}, \beta_r = 5 J_l^{-1}$, $\Delta_l=-\Delta_r=-0.3$, $J_r/J_l=1$.}
\label{fig:theta_evolution_2}
\end{figure}
%
%%%%%%%%%%%%%%%%%%%%%%%%%%%%%%%%%%%%%%%%%%%%%%%%%%%%%%%%%%%%%%%%% 

We are now in the position to integrate the previous results into a general framework for the equilibration dynamics of inhomogeneous chains consisting of two different integrable models.
We remark that our problem is a particular kind of local quantum quench, where the initial state differs from a steady state of the evolving Hamiltonian only at the 
junction. 
The local unitary dynamics induced by $\hat H$ abides by a 
Lieb-Robinson bound, namely by a spreading of correlations which is constrained to be within a light cone with typical velocity $v$, propagating from the center. 
Outside the light-cone, the state at time $t$ is 
indistinguishable from the initial one apart from exponentially small corrections. 
Generally speaking, the aim of this section is to describe the dynamics within the light-cone both for the time regimes accessible with our numerics and for the asymptotical $t \to \infty$ stationary properties.

\subsection{Pre-equilibration behaviour}
\label{sec:preequilibration}

Let us first consider the time regime which is accessible with our numerics. The results in Figs.~\ref{fig:XXZ_scalingT}, \ref{fig:derivativeG} and~\ref{fig:ramp} show that after a very short transient behaviour of order $t \sim J_l^{-1}$, the energy current $\mathcal J(t)$ is quasi-stationary, with relative variations which are negligible within the considered time window.

In Fig.~\ref{fig:theta_evolution_2} (top) we further elaborate on this point and show the current profile $\langle \hat j_{\lambda,n} \rangle_t$ (where $\braket{\ldots}_t$ is the expectation value over the state at time $t$) for several values of $t$. The propagation is asymmetric due to the different properties of $\hat H_l$ and $\hat H_r$. Moreover, the current around the junction rapidly becomes homogeneous and quasi-stationary. This is explicitly verified in Fig.~\ref{fig:theta_evolution_2} (bottom), where the averaged current over $11$ sites approaches $\mathcal J(t)$, while both become almost time-independent.

%%%%%%%%%%%%%%%%%%%%%%%%%%%%%%%%%%%%%%%%%%%
%
\begin{figure}[t]
\includegraphics[width=\columnwidth]{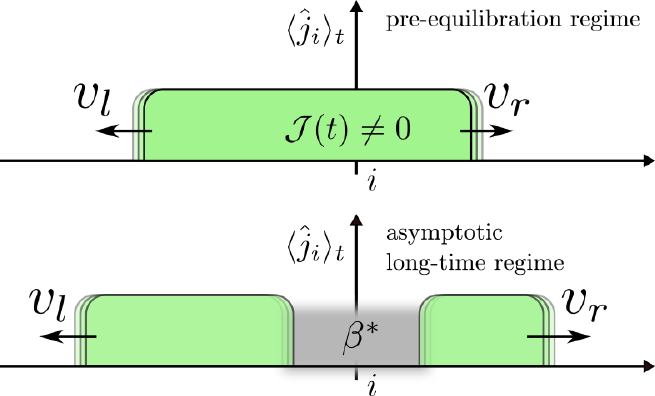}
\caption{(color online) Top panel: sketch of the energy current profile in the pre-equilibration behaviour
as described by the low-energy field theory: two {wavefronts} propagate in opposite direction with different 
velocities $v_l, v_r$ determining an almost stationary and homogeneous energy current around the junction.
Lower panel: sketch of the asymptotic long-time regime; a thermal region at inverse temperature $\beta^\ast$
develops around the junction due to the
local integrability breaking. The two transient regions of non-vanishing current keep travelling and increasing their size,
as discussed in Sec.~\ref{sec:asymptotic}.}
\label{fig:sketch:3}
\end{figure}
%
%%%%%%%%%%%%%%%%%%%%%%%%%%%%%%%%%%%%%%%%%%%

In between this central region and the unperturbed areas outside the light cone, two transient zones appear: the current here is not $\mathcal J(t)$ nor zero, as at the initial time. 
Given a time $t$ we select $b,-a \gg vt$; if we neglect these transient regions, it is possible to draw a connection between $\mathcal J(t)$ and $\langle \hat j_{a,b} \rangle_t$ defined in Eq.~\eqref{eq:jabdef}. Indeed, because of the homogeneity of the current profile discussed above, it is possible to make the following approximation: $\langle \hat j_{a,b} \rangle_t \sim  \mathcal J(t) (v_l+v_r) t$; since $\mathcal J(t)$ is quasi-stationary the following holds:
\begin{equation}
 \mathcal J(t) \sim \frac{\langle 
 \partial_t \hat j_{a,b} \rangle_t}{v_l+v_r} .
 \label{eq:fondamentale2}
\end{equation}
Interestingly, it is possible to give a more explicit expression to $\langle 
 \partial_t \hat j_{a,b} \rangle_t$ using
Eq.~\eqref{eq:Theta} and the Lieb-Robinson bound, which lead to the following formula:
\begin{equation}
 \langle \partial_t \hat j_{a,b} \rangle_t = 
 \braket{\hat k_{l,a}}_{0}-\braket{\hat k_{r,b}}_{0} + 
\braket{\hat\Theta}_t.
\label{eq:fondamentale}
\end{equation}
We remark that Eq.~\eqref{eq:fondamentale} is true up to exponentially small corrections.

\begin{figure}[t]
\centering
\includegraphics[width=0.40\textwidth]{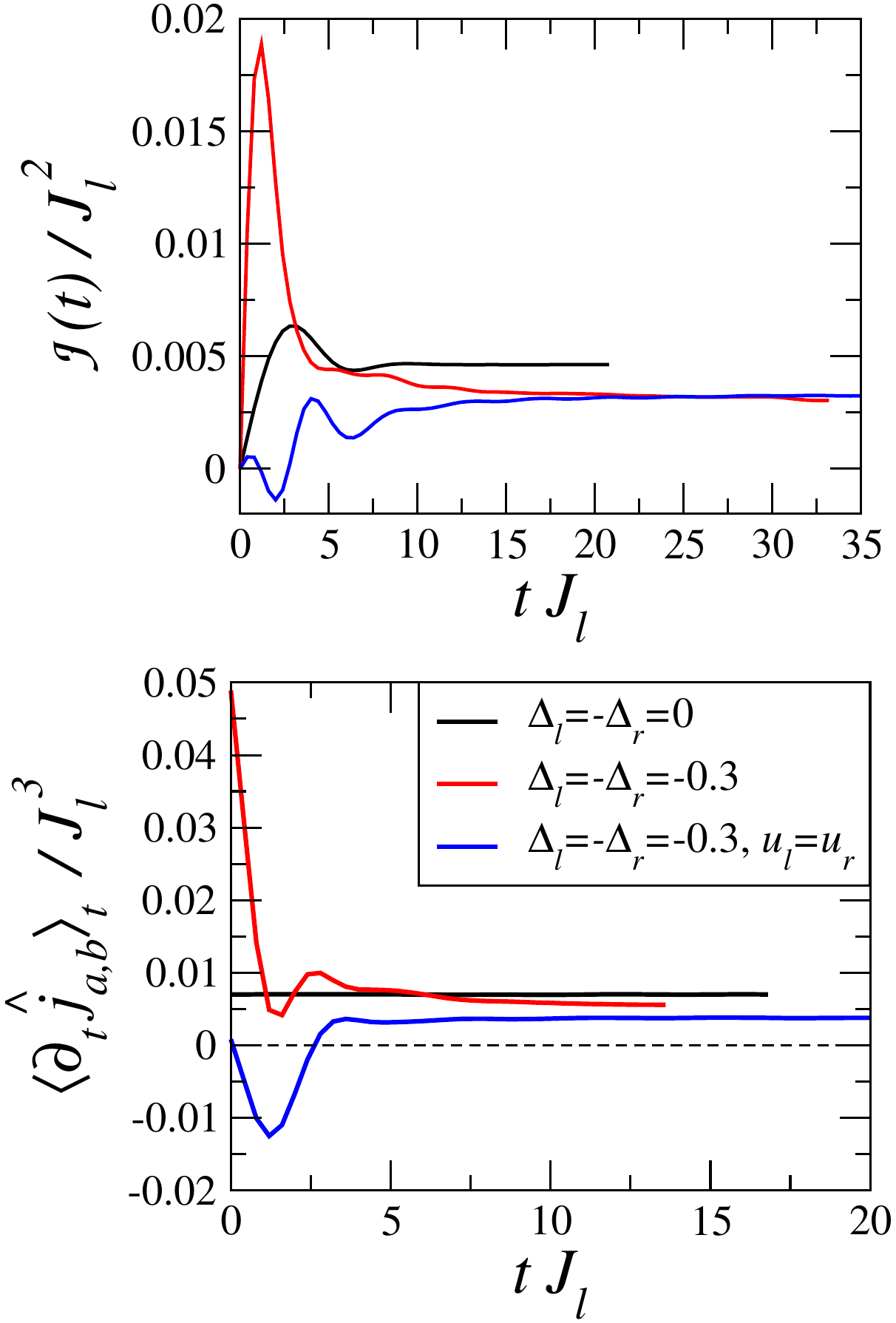}
\caption{(color online) Top panel: time evolution of the energy current $\mathcal J(t)$ at the center of the chain.
Lower panel: time evolution of the derivative of the integrated energy current $\braket{\partial_t\hat J_{ab}}_t$.
We considered the homogeneous case (black lines), 
inhomogeneous case with $u_l\neq u_r$ (red lines) and inhomogeneous case with 
$u_l= u_r$ (blue lines).
We set $\beta_l=4 \ J_l^{-1}$, $\beta_r=5 \ J_l^{-1}$, $\Delta_l=-\Delta_r=-0.3$ 
corresponding to 
$u_l=0.798(9) J_l$ and $u_r=1.183(5) J_l$.
The hopping parameter is $J_r/J_l=1$ apart from the velocity matching case where 
it is chosen so that $u_l=u_r$ is satisfied. 
}
  \label{fig:kabtheta}
\end{figure}

In Fig.~\ref{fig:kabtheta} we plot $\mathcal J(t)$ and $\langle \partial_t \hat j_{a,b} \rangle_t$: the comparison shows interesting analogies which seem to hold even though the transient regions, plotted in Fig.~\ref{fig:theta_evolution_2} (top), are significant. In general, Eq.~\eqref{eq:fondamentale2} is well justified whenever the transient regions grow as $t^{\alpha}$ with $0 \leq \alpha < 1$, introducing only subleading corrections (see for example the case discussed in Ref.~\cite{doyon_2015_B}, but note that for free fermions this is not verified~\cite{DW2015}).

%%%%%%%%%%%%%%%%%%%%%%%%%%%%%%%%%%%%%%%%%%s%%%%%%%%%%%%%%%%%%%%%%% 
%
%%%%%%%%%%%%%%%%%%%%%%%%%%%%%%%%%%%%%%%%%%%%%%%%%%%%%%%%%%%%%%%%% 

Let us compare these results with those presented in the literature for homogeneous integrable systems~\cite{bernard2012, bernard2015}.
If $\Delta_l = \Delta_r$ and $J_l=J_r$~\cite{deluca2014}, the operator 
$\hat\Theta=0$ and thus $\braket{\hat j_{a,b}}_t\propto t$, with slope given by $\braket{\hat k_{l,a}}_{0}-\braket{\hat k_{r,b}}_{0}$. 
This slope can be shown to give an exact lower bound to the stationary energy current at the junction $\mathcal J$~\cite{Doyon2015}.
This simple case corresponds to a perfectly transmitting junction, as expected because the system is homogeneous.

In general, the behaviour of $\braket{\hat\Theta}_t$ allows us to discern among more complicated scenarios in which transmission might not be perfect. In particular, whenever $\langle \hat \Theta \rangle_t $ is quasi-stationary in time and different from $\braket{\hat k_{l,a}}_{0}-\braket{\hat k_{r,b}}_{0}$, it is intriguing to interpret it as an impurity determining a finite transmission because the matching between the modes of the halves is not perfect~\cite{Viti2015}.
This is for instance what is discussed in Sec.~\ref{sec:low_energy}, where the inhomogeneous low-energy theory of the system introduces a transmission coefficient $\mathcal T$ and thus a non-optimal transmission of energy.

\subsection{Asymptotic long-time regime \label{sec:asymptotic}}

Let us now move to the study of the system in the asymptotic long-time regime which is not accessible with our numerics. 
{Although nobody has yet identified the minimal conditions for thermalisation, the expectation is that generic systems (e.g. exhibiting Wigner-Dyson level spacing statistics) should thermalize.}
We have argued that at long times the current at the junction $\mathcal J(t)$ decays to zero (see Sec.~\ref{sec:problem}). 
Whereas this steady-state property is a direct consequence of the assumption of thermalisation, it is possible to further elaborate on how the system approaches this situation.
In particular, we will now elaborate on the existence of two domains of growing size which
(i) are characterized by a non-zero energy current and (ii) propagate from the junction towards the two extremities of the chain (see the sketch in Fig.~\ref{fig:sketch:3}). 

We begin by showing that $\langle \hat j_{a,a'} \rangle_t$ must grow linearly in time for $a,a'<0$ and $-a \gg vt \gg -a'$. Starting from the definition, the time-derivative reads:
\begin{equation}
 \langle \partial_t \hat j_{a,a'} \rangle_t =
 \braket{\hat k_{l,a}}_0- 
 \braket{\hat k_{l,a'}}_t .
\end{equation}
In this equation, by choosing $a$ properly, $t$ can be taken large enough to ensure $\braket{\hat k_{l,a'}}_t$ to have stationary and thermal behaviour.
Assuming to be at times $t$ so long that thermalisation in the region around $a'$ has occurred, 
\begin{equation}
\braket{\hat k_{l,a'}}_t = \frac{ \text{Tr}[\hat k_{l,a'} e^{-\beta^* \hat H}] }{ \text{Tr}[ e^{-\beta^* \hat H}]}.
\end{equation}
We postpone the detailed analysis on the exact value of $\beta^*$ to Sec.~\ref{subsec:effbeta}, here it suffices to assume the reasonable inequality $\beta_l < \beta^* < \beta_r$. 
Recalling that $a$ is still outside the light cone and thus $\braket{\hat k_{l,a}}_0$ is a thermal expectation value with inverse temperature $\beta_l$, the following holds:
\begin{equation}
\braket{\hat k_{l,a}}_0- \braket{\hat k_{l,a'}}_t>0 
\end{equation}
and the integrated current $\langle \hat j_{a,a'} \rangle_t$ increases linearly in time as long as the wavefront has not reached $a$.
A simple analysis shows that this happens also on the right half of the system in a similarly defined $b'$-$b$ region.

How is it possible to make this result compatible with the fact that for long times $\mathcal J(t) \to 0$? 
The simplest scenario entails two domains which propagate in opposite directions, respectively in the right and in the left half. 
The size of each domain naturally grows because its two edges propagate due to different physical phenomena. 
The most external edge moves with the wavefront velocity at which energy is spreading in the chain. 
The most internal one is related to the spread of thermalisation originating from the junction.
Thus, even if at short times the quenched dynamics resembles that of the homogeneous case, as discussed in Sec.~\ref{sec:preequilibration}, the integrability-breaking induces thermalisation and accordingly a slow decay to zero of $\mathcal J(t)$. This feature first appears at the junction and then affects all the other sites of chain.

A last comment is in order. Since
$\braket{\hat j_{a,b}}_t =
\braket{\hat j_{a,a'}}_t+
\braket{\hat j_{a',b'}}_t+
\braket{\hat j_{b',b}}_t$, 
using the previous results we obtain that 
$\braket{\hat j_{a,b}}_t$ also grows 
linearly in time. Indeed, $\hat j_{a',b'}$ is a bounded operator that cannot compensate the linear growth in time of the other two terms.
Since it has been shown that $\braket{\hat j_{a,b}}_t$ grows linearly in time also for the homogeneous chain, where $\mathcal J \neq 0$, this indicates 
that there is no general connection between $\mathcal J$ and $\braket{ \partial_t \hat j_{a,b}}_t$ in the asymptotic long-time limit.
The discussion related to Eq.~\eqref{eq:fondamentale2} is valid only for intermediate times where quasi-stationarity appears and $\mathcal J(t)$ is still different from zero.

\subsection{The problem of the effective temperature}
\label{subsec:effbeta}

We now consider the problem of identifying $\beta^*$. 
{
In highly inhomogeneous scenarios, such as the one considered in this article, different regions of the system may experience thermalisation both (i) at different times and (ii) at different effective temperatures.
Here, for example, the asymptotic extrema of the chain are always thermal at their initial temperatures; additionally, we expect thermalisation at $\beta^*$ to occur first at the junction and then propagate towards the edges. 
In similar situations with well-defined initial temperatures and unitary dynamics, the problem has been approached within the framework on Onsager kinetic equation~\cite{Georges2013}.
In order to link these methods to our microscopic model, it would be necessary to derive, at a coarse-grained level, a hydrodynamical equation governing the temperature dynamics in our system. Here we limit ourselves to some general considerations related to some fundamental constraints for a microscopic theory of thermalisation in a inhomogeneous system.}

{In typical quench problems,}
the conservation of energy (the time evolution is unitary) makes it natural to define {$\beta^*$} through the following implicit equation:
\begin{equation}
 \frac{\text{Tr}[\hat H_l e^{- \beta_l \hat H_l}]}{\text{Tr}[e^{- \beta_l \hat H_l}]} + 
 \frac{\text{Tr}[\hat H_r e^{- \beta_r \hat H_r}]}{\text{Tr}[e^{- \beta_r \hat H_r}]} 
 =
 \frac{\text{Tr}[\hat H e^{- \beta^* \hat H}]}{\text{Tr}[e^{- \beta^* \hat H}]}.
 \label{eq:beta:star}
\end{equation}
However, if one considers a situation where the ratio between the lengths of the left and right halves is not $1$, as assumed until now, but a different fraction, one realizes that the estimation of $\beta^*$ from Eq.~\eqref{eq:beta:star} changes. This is absurd as the local dynamics 
does not involve the boundary and thus is not affected by this rescaling.
Intuitively, this equation assumes that the stationary state inside the light cone is affected by the total amount of energy, and thus also by the fraction which lies outside it. However, this latter fraction could not play any role in the equilibration dynamics and cannot affect the value of the stationary temperature.
{This discussion highlights the fact that thermalisation is a local process which only in fine-tuned situations (e.g. completely homogeneous systems) can be addressed via global conserved quantities like Eq.~\eqref{eq:beta:star}.}

From a practical point of view, one can envision to define $\beta^*$ through the long-time dynamics of a local observable sitting at the junction. If the stationary state is thermal-like, all local observables should agree with the thermal expectation value in the long-time regime.

\section{Conclusions}
In this work we studied energy transport between two (ideally) semi-infinite XXZ chains characterised by different coupling constants. The {\it partitioning} protocol 
considered here is identical to the one discussed in several papers recently appeared in the literature: the two half-chains are initially prepared in two separate 
thermal states and kept at different temperature. The chains are then connected, the system is let unitarily evolve and the energy current flowing 
through the system is analysed as a function of time. 

Most of the paper is devoted to the case where each half-chain has uniform couplings; there is, however, a difference in their value between the left and right halves. Each half-chain 
is integrable, the whole system is not. This setup allowed us to study in details the scattering mechanisms regulating the energy flow in the system, 
as well as their relation to the integrability of the underlying model and to its conservation laws. In order to dwell more into the scattering of bosonic modes and quasi-particles
at the junction, we also considered a situation in which the couplings are (adiabatically) varied through a  finite region.
The results that we presented are obtained both with a bosonisation approach and with numerical simulations based on time-dependent 
matrix product states. 

The general scenario that emerges is quite rich including a pre-thermalisation regime and a final thermalisation. Since integrability is only 
locally broken in this model, a pre-equilibration behaviour may appear. In particular, when the sound velocities of the bosonic modes 
of the two halves match, the low-temperature energy current is almost stationary and is described by a formula with a non-universal 
prefactor interpreted as a transmission coefficient. Thermalisation, characterised by the absence of any energy flow, occurs only on 
longer time-scales which are not accessible with our numerics.

Quite interesting in this respect would be the investigation of the transition between the pre-equilibration and the asymptotic long-time regime: What is the typical time-scale of the thermalisation onset? How does it depend on the system parameters and on the initial temperatures? Another attractive direction would be the development of an analytical approach to this problem at high temperatures where the results obtained exploiting the low-energy field-theory are not reliable.
{It is not to be excluded that also the investigation of higher-order cumulants of the full counting statistics may provide valuable information~\cite{Saleur2014}.}

\acknowledgments

We acknowledge enlightening discussions with: J. Dubail, M. Haque, J. E. Moore, J-.M. Stephan  and X. Zotos.
AB, DR, LM, and RF acknowledge the Kavli Institute for Theoretical Physics, University of California, Santa Barbara (USA) 
for the hospitality and support during the completion of this work.
This work was supported in part by the National Science Foundation under Grant No. NSF PHY11-25915, by “Investissements d’Avenir” LabEX PALM (ANR-10-LABX-0039-PALM) (AD), by
LabEX ENS-ICFP: ANR-10-LABX-0010/ANR-10-IDEX-0001-02 PSL* (LM),  by EU (IP-SIQS) and (STREP-QUIC) (RF), 
by the Italian MIUR (FIRB project RBFR12NLNA) (DR), and by Scuola Normale Superiore through the internal project ``Non-equilibrium dynamics of one-dimensional quantum systems'' (RF, DR, and LM).

%%%%%%%%APPENDIX%%%%%%%%%%

\appendix

\section{Level spacing statistics}
\label{app:levstat}

In this section we study how the level spacing statistics (LSS) of the XXZ chain is affected by the inhomogeneity discussed in the main text.
We perform an exact diagonalization of systems made of $N = 16$ spins in the zero magnetization sector ($12870$ states), and we analyzed the distribution $P(s)$ of the normalized level spacings $s_n=(E_{n+1}-E_n)/D$, where $E_n$ are the eigenvalues of~\eqref{eq:Hamiltonian} in ascending order (we discarded the lowest and the higher $10^3$), and $D$ is the average level spacing.

%%%%%%%%%%%%%%%%%%%%%%%%%%%%%%%%%%%%%%%%%%%%%%%%%%%%%%%%%%%%%%%%% 
%
\begin{figure}[b]
\centering
\includegraphics[width=0.42\textwidth]{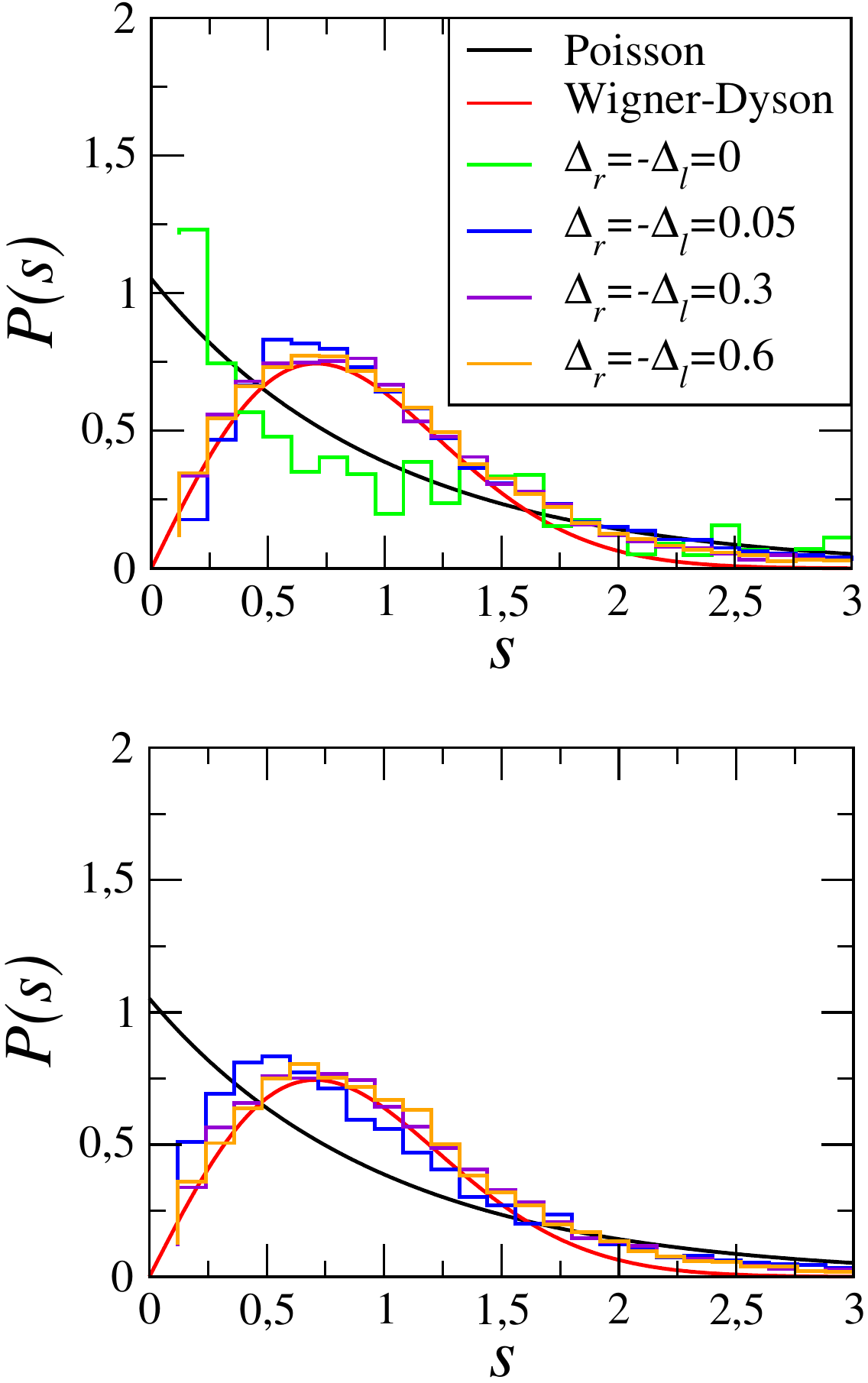}
\caption{(color online) Distribution of the normalized level spacings for a chain made of $N=16$ spins in the zero magnetization sector.
Top panel: homogeneous ($\Delta_l=\Delta_r=0$) and inhomogeneous cases with $u_l\neq u_r$ (choosing $\Delta_r=-\Delta_l>0$). The hopping parameters are fixed as $J_r/J_l=1$.
Lower panel: inhomogeneous case with $\Delta_{l}$ and $\Delta_{r}$ fixed as in the top panel but tuning $J_r/J_l$ in order to get $u_l=u_r$. 
 }
  \label{fig:histo_L16}
\end{figure}
%
%%%%%%%%%%%%%%%%%%%%%%%%%%%%%%%%%%%%%%%%%%%%%%%%%%%%%%%%%%%%%%%%%

In the top panel of Fig.~\ref{fig:histo_L16} we compare the results for the homogeneous case ($\Delta_l=\Delta_r=0$ and $J_r/J_l=1$) with the inhomogeneous case with $u_l\neq u_r$ choosing $\Delta_r=-\Delta_l>0$ and $J_r/J_l=1$ (which corresponds to the $K=1$ case).
From our data we observe that the level spacing distribution $P(s)$ moves from a Poissonian shape, $P_{\rm P}(s)\propto{\rm e}^{-s}$, to a Wigner-Dyson distribution, $P_{\rm WD}(s)\propto s \ {\rm e}^{-\pi s^2/4}$, very quickly as soon as we switch on the anisotropy term $\Delta_r=-\Delta_l \neq 0$.
A Wigner-Dyson distribution of the LSS clearly signals the tendency toward level repulsion~\cite{haake}, a typical feature of non-integrable systems which seems to be present in this context at any value of the anisotropy strength $\Delta_r=-\Delta_l$.

\begin{figure}[t]
\centering
\includegraphics[width=0.42\textwidth]{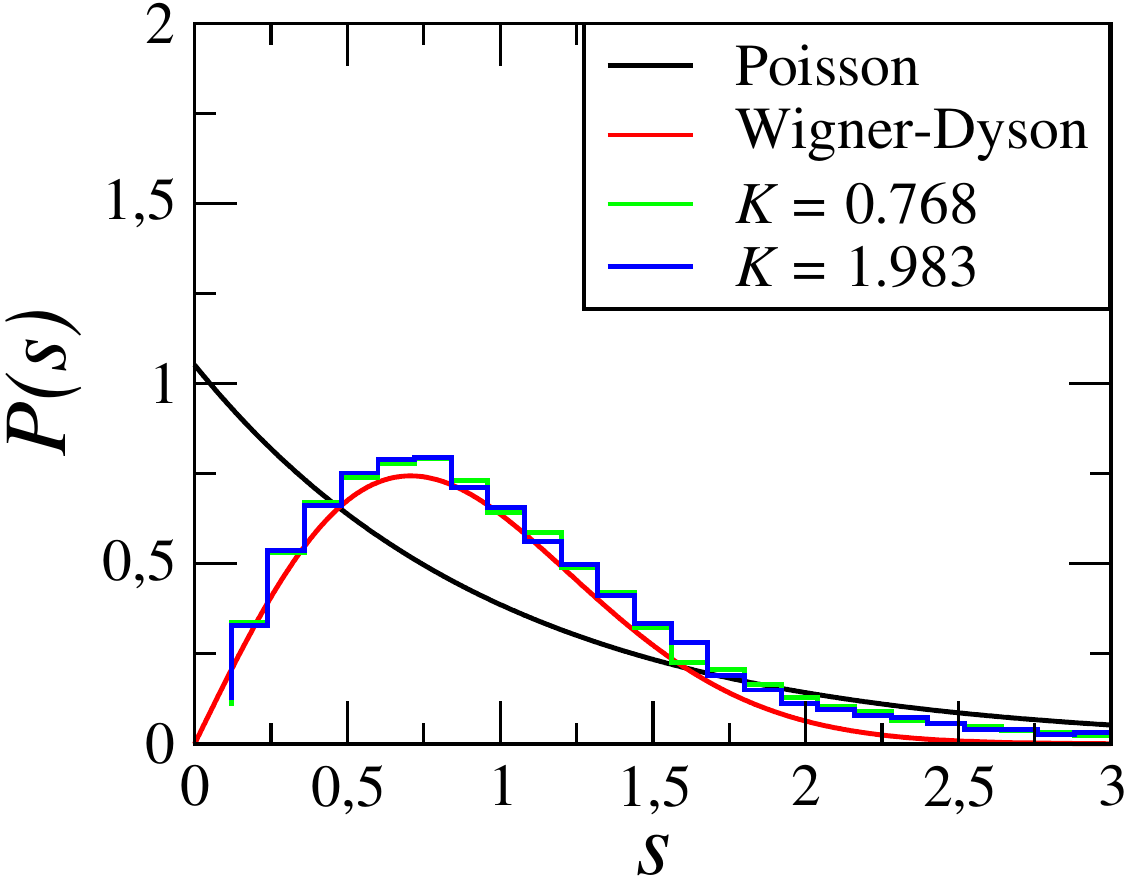}
\caption{(color online) Distribution of the normalized level spacings for a chain made of $N=16$ spins in the zero magnetization sector.
Inhomogeneous cases with $u_l\neq u_r$ choosing $\Delta_l=1$ and $\Delta_r=0.95$ (green line), $-0.95$ (blue line). The hopping parameters are fixed as $J_r/J_l=1$.
 As indicated in the legend, the different values of $\Delta_r$ allow to explore the $K \neq 1$ cases.
 }
  \label{fig:histo_L16_2}
\end{figure}

In the lower panel of Fig.~\ref{fig:histo_L16} we perform the same analysis for the same anisotropy parameters but tuning $J_r/J_l$ in order to get $u_l=u_r$ and we observe that the integrability is broken in the same way.
In order to explore also the $K\neq1$ cases we set the anisotropy parameters as in Fig.~\ref{fig:derivativeG}.
The results shown in Fig~\ref{fig:histo_L16_2} indicate, also in these cases, a clear tendency toward a Wigner-Dyson distribution.

In conclusion, our LSS data indicate that in our system~\eqref{eq:Hamiltonian} the integrability is always broken in the inhomogeneous case $\Delta_l\neq\Delta_r$, regardless of the value of $K$ and whether the renormalized Fermi velocities $u_{l,r}$ are matched or not.

%%%%%%%%%%%%%%%%%%%%%%%%%%%%%%%%%%%%%%%%%%%%%%%%%%%%%%%%%%%%%%%%% 
%
%
%%%%%%%%%%%%%%%%%%%%%%%%%%%%%%%%%%%%%%%%%%%%%%%%%%%%%%%%%%%%%%%%%  

\section{Details on the $\hat \Theta$ operator}
\label{app:theta}

In this Appendix we provide the explicit form of the operators $\hat k_{\lambda,n}$ and $\hat \Theta$ appearing in Eq.~\eqref{eq:Theta}.
Let us consider the {\it integrated} energy current operator $\hat j_{a,b}$ as defined in Eq.~\eqref{eq:jabdef}.
% \begin{equation}
% \hat j_{a,b} = \sum_{n=a}^{-1} \hat j_{l,n} +
%  2\hat j_0 +
%  \sum_{n=2}^b \hat j_{r,n},
%  \qquad a<0, \; b >0.
%  \label{eq:jabdef_app}
% \end{equation}
By computing the time derivative of such operator one gets Eq.~\eqref{eq:Theta}, which is here reproduced for the sake of readability:
\begin{equation}
i [\hat H, \hat j_{a,b}] = \hat k_{l,a} - \hat k_{r,b} + \hat \Theta.
\end{equation}
The operators $\hat k_{l,a}$ and $\hat k_{r,b}$ are linear combinations of two- and four-point operators with support on few sites around $a$ and $b$ respectively; 
$\hat\Theta$ is the sum of two-point operators localized around the junction.

Using the definition~\eqref{eq:energy:current:density} of the {\it local} energy current $\hat j_{\lambda,n}$ and exploiting the canonical commutation relations $\bigl[ \hat S_n^\alpha, \hat S_m^\beta  \bigr]={\rm i} \delta_{n m} \epsilon_{\alpha\beta\gamma} \hat S_n^\gamma$ (where $\delta_{nm}$ is the Kronecker delta and $\epsilon_{\alpha\beta\gamma}$ is the Levi-Civita symbol), after some algebra we get:
\begin{widetext}
\begin{eqnarray}
\hat k_{l,a} &=&
\frac{J_l^3}{4}  \Bigl[ 
\bigl( \Delta_l^2+1 \bigr) 
\bigl( \hat S_{a-1}^x  \hat S_a^x +
\hat S_{a-1}^y \hat S_a^y \bigr)
+2  \Delta_l  \Bigl(2  \bigl(- \Delta_l  (\hat S_{a-2}^z  \hat S_{a-1}^x  \hat S_{a}^x  \hat S_{a+1}^z + \hat S_{a-2}^z \hat S_{a-1}^y  \hat S_a^y  \hat S_{a+1}^z  ) 
+ \hat S_{a-2}^x  \hat S_{a-1}^y \hat S_a^x \hat S_{a+1}^y + 
\nonumber \\
&& - \hat S_{a-2}^x  \hat S_{a-1}^y \hat S_a^y  \hat S_{a+1}^x 
- \hat S_{a-2}^y \hat S_{a-1}^x \hat S_a^x \hat S_{a+1}^y + \hat S_{a-2}^y  \hat S_{a-1}^x  \hat S_a^y  \hat  S_{a+1}^x +\hat S_{a-2}^x  \hat S_{a-1}^z  \hat S_a^x \hat S_{a+1}^z +\hat S_{a-2}^z \hat S_{a-1}^x \hat S_a^z  \hat S_{a+1}^x+  \nonumber \\
&& +\hat S_{a-2}^y \hat S_{a-1}^z \hat S_a^y \hat S_{a+1}^z + \hat S_{a-2}^z \hat S_{a-1}^y \hat S_a^z \hat S_{a+1}^y  \bigr)+ \hat S_{a-1}^z  \hat S_a^z  \Bigr)
-4  \bigl(\hat S_{a-2}^x \hat S_{a-1}^z \hat S_a^z \hat S_{a+1}^x+\hat S_{a-2}^y \hat S_{a-1}^z \hat S_a^z \hat S_{a+1}^y  \bigr)
\Bigr]; \\
&& \nonumber \\
\hat k_{r,b} &=&
\frac{J_r^3}{4}  \Bigl[ \bigl( \Delta_r^2+1 \bigr) 
\bigl( \hat S_{b}^x  \hat S_{b+1}^x 
+\hat S_{b}^y \hat S_{b+1}^y  \bigr)
+2  \Delta_r  \Bigl(2  \bigl(- \Delta_r  (\hat S_{b-1}^z  \hat S_{b}^x  \hat S_{b+1}^x  \hat S_{b+2}^z + \hat S_{b-1}^z \hat S_{b}^y  \hat S_{b+1}^y  \hat S_{b+2}^z  ) + \hat S_{b-1}^x  \hat S_{b}^y \hat S_{b+1}^x \hat S_{b+2}^y + \nonumber \\
&&- \hat S_{b-1}^x  \hat S_{b}^y \hat S_{b+1}^y  \hat S_{b+2}^x 
- \hat S_{b-1}^y \hat S_{b}^x \hat S_{b+1}^x \hat S_{b+2}^y + \hat S_{b-1}^y  \hat S_{b}^x  \hat S_{b+1}^y  \hat  S_{b+2}^x +\hat S_{b-1}^x  \hat S_{b}^z  \hat S_{b+1}^x \hat S_{b+2}^z +\hat S_{b-1}^z \hat S_{b}^x \hat S_{b+1}^z  \hat S_{b+2}^x+ \nonumber \\
&& +\hat S_{b-1}^y \hat S_{b}^z \hat S_{b+1}^y \hat S_{b+2}^z + \hat S_{b-1}^z \hat S_{b}^y \hat S_{b+1}^z \hat S_{b+2}^y  \bigr)+ \hat S_{b}^z  \hat S_{b+1}^z  \Bigr)
-4  \bigl(\hat S_{b-1}^x \hat S_{b}^z \hat S_{b+1}^z \hat S_{b+2}^x+\hat S_{b-1}^y \hat S_{b}^z \hat S_{b+1}^z \hat S_{b+2}^y  \bigr)
\Bigr]; \\
&& \nonumber \\
\hat \Theta &=& \frac{1}{4}  \Bigl[ \hat S_{1}^x   \hat S_{2}^x  \Bigl( \bigl( \Delta_r^2+1 \bigr)  J_r^3- \bigl( \Delta_l^2+1 \bigr)  J_l^2  J_r \Bigr)+\hat S_{1}^y  \hat S_{2}^y  \Bigl( \bigl( \Delta_r^2+1 \bigr)  J_r^3- \bigl( \Delta_l^2+1 \bigr)  J_l^2  J_r \Bigr)+ J_l \hat S_{0}^x   \hat S_{1}^x  \Bigl( \bigl( \Delta_r^2+1 \bigr)  J_r^2+ \nonumber \\
&&- \bigl( \Delta_l^2+1 \bigr)  J_l^2 \Bigr)+ J_l \hat S_{0}^y  \hat S_{1}^y  \Bigl( \bigl( \Delta_r^2+1 \bigr)  J_r^2- \bigl( \Delta_l^2+1 \bigr)  J_l^2 \Bigr)+2  \Delta_l  J_l  \bigl( J_r^2- J_l^2 \bigr) \hat S_{0}^z  \hat S_{1}^z+2  J_l  J_r \hat S_{0}^x   \hat S_{2}^x \bigl( \Delta_l  J_l- \Delta_r  J_r \bigr)+ \nonumber \\
&& +2  J_l  J_r \hat S_{0}^y  \hat S_{2}^y \bigl( \Delta_l  J_l- \Delta_r  J_r \bigr)+2  J_l  J_r \hat S_{0}^z  \hat S_{2}^z \bigl( \Delta_r  J_l- \Delta_l  J_r \bigr)+2  \Delta_r  J_r \bigl( J_r- J_l \bigr) \bigl( J_l+ J_r\bigr) \hat S_{1}^z  \hat S_{2}^z \Bigr]. 
\end{eqnarray}
\end{widetext}
The cumbersome expression for $\hat\Theta$ is significantly simplified when $J_l=J_r$:
\begin{align}
\hat \Theta = &
-\frac{J_r^3}{4}  \bigl(\Delta_l-\Delta_r\bigr) \Bigl[\bigl(\Delta_l+\Delta_r\bigr) \Bigl(\hat S_{0}^x \hat S_{1}^x+\hat S_{1}^x \hat S_{2}^x+ \nonumber \\
&+ \hat S_{0}^y \hat S_{1}^y+ \hat S_{1}^y \hat S_{2}^y\Bigr)-2 \Bigl(\hat S_{0}^x \hat S_{2}^x+\hat S_{0}^y \hat S_{2}^y-\hat S_{0}^z \hat S_{2}^z \Bigr)\Bigr]. 
\end{align}
Additionally, when $J_l=J_r$ and $\Delta_l=\Delta_r$ (homogeneous case) we obtain $\hat\Theta=0$.

\section{Details on analytical calculations}
\label{app:CFT}

Here we detail the computation leading to the result~\eqref{eq:current_CFT}. The equation~\eqref{eq:LR_t} is the
action of the left-to-right transfer matrix on the fields $\partial_x\phi_l$ and $\partial_t\phi_l$. In order to have the
same normalization 
for the stress-energy tensor on the left and on the right one has to redefine~\cite{Bachas} the bosonic fields as 
\begin{equation}
\label{varphi}
\phi(x)=\left\{
\begin{array}{l}
\sqrt{K_l}\varphi(x)\quad x<0\\
\sqrt{K_r}\varphi(x)\quad x\geq 0
           \end{array}\right.
\end{equation}
Further, introducing the notation $\varphi_{r/l}\equiv\varphi(0^{\pm})$, we can rewrite \eqref{eq:LR_t} as
\begin{equation}
 \label{TM}
 \left[
\begin{array}{l}
\partial_x\varphi_l\\

\partial_t\varphi_l \end{array}\right]=\left[
\begin{array}{ll}
\alpha^{1/2} & 0\\
0 & \alpha^{-1/2}
                                    \end{array}\right]\left[
\begin{array}{l}
\partial_x\varphi_r\\

\partial_t\varphi_r \end{array}\right].
\end{equation}
Notice that the equation above ensures the continuity of the  momentum density through the junction
$p_l(x,t)=p_r(x,t)$, with $p_{l/r}=\partial_x\varphi_{l/r}
\partial_t\varphi_{l/r}$. Total momentum is however not conserved by the dynamics.

\begin{figure}[t]
\centering
\begin{tikzpicture}[scale=1.5]
\draw[thick,->](-1,-1)--(-0.5,-0.5);
\draw[thick](-0.5,-0.5)--(0,0);
\draw[thick,->](1,-1)--(0.5,-0.5);
\draw[thick](0.5,-0.5)--(0,0);
\draw[thick](-0.5+1,-0.5+1)--(0+1,0+1);
\draw[thick,->](-1+1,-1+1)--(-0.5+1,-0.5+1);
\draw[thick,->](1-1,-1+1)--(0.5-1,-0.5+1);
\draw[thick](0.5-1,-0.5+1)--(0-1,0+1);
\draw[yellow, fill=yellow!50] (-0.1,-1) rectangle (0.1,1);
\draw[->](-2,-1)--(2,-1);
\draw(2,-1) node[above] {$x$};
\draw[->](-2,-1)--(-2,-0.5);
\draw(-2,-0.5) node[left] {$t$};
\draw(-1,-1) node[below]{$\bar{\partial}\varphi_l$};
\draw(1,-1) node[below]{$\partial\varphi_r$};
\draw(-1,1) node[above]{$\partial\varphi_l$};
\draw(1,1) node[above]{$\bar{\partial}\varphi_r$};
\draw(-0.1,0) node[left]{$\Omega$};
\end{tikzpicture}
\caption{(color online) The dual defect map $\Omega$ acting on the incoming and outgoing chiral bosonic waves.}
\label{figscattering}
\end{figure}
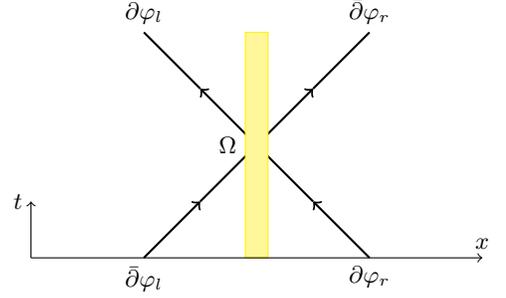

To connect with the CFT formalism we must introduce the
chiral waves $\partial \varphi=\frac{1}{2}[\partial_x\varphi-\partial_t\varphi]$ and
$\bar{\partial} \varphi=\frac{1}{2}[\partial_x\varphi+\partial_t\varphi]$ and recast \eqref{TM} in the form a scattering matrix,
see Fig. \ref{figscattering}. 

One finds
 \begin{equation}
 \label{theta_op}
 \left[
\begin{array}{l}
\partial\varphi_r\\

\bar{\partial}\varphi_l \end{array}\right]=\left[
\begin{array}{ll}
\cos\gamma & \sin\gamma\\
-\sin\gamma & \cos\gamma
                                    \end{array}\right]\left[
\begin{array}{l}
\partial\varphi_l\\

\bar{\partial}\varphi_r \end{array}\right].
\end{equation}
with $\cos\gamma=\displaystyle\frac{2\sqrt{\alpha}}{1+\alpha}$. The matrix in \eqref{theta_op} can be interpreted as the
representation of the action of the dual defect map $\Omega$ on the incoming waves~\cite{Viti2015}.
To compute the energy current one needs to apply the composition of maps $\Omega_0^{-1}\circ\Omega$ to the
momentum operator $T-\bar{T}$ and evaluate the result on the initial state $\rho_0=\rho_l\otimes\rho_r$, where the two
free bosonic CFTs are in a thermal Gibbs ensemble at inverse temperature $\beta_{l/r}$. We recall that the
stress-energy tensor components are $T=(\partial\varphi)^2$ and $\bar{T}=(\bar{\partial}\varphi)^2$ and that the map $\Omega_0$ acts as a pure
reflection on the chiral incoming waves
\begin{equation}
 \label{theta_not}
 \Omega_0\left[
\begin{array}{l}
\partial\varphi_r\\

\bar{\partial}\varphi_l \end{array}\right]=\left[
\begin{array}{ll}
0 & 1\\
-1 & 0
                                    \end{array}\right]\left[
\begin{array}{l}
\partial\varphi_l\\

\bar{\partial}\varphi_r \end{array}\right].
\end{equation}
Since we assumed the asymptotic value of the current $x$-independent (see however the discussion at the end of Sec. \ref{sec:low_energy}) we can can take $x=0^-$ and obtain by simple manipulation
$\Omega_0^{-1}\circ\Omega[T_l]=\bar{T}_l$ and 
\begin{equation}
\label{thetabarT}
\Omega_0^{-1}\circ\Omega[\bar{T}_l]=
\cos^2\gamma~T_r+\sin^2\gamma~\bar{T}_l+2\cos\gamma\sin\gamma~\partial\phi_r\bar{\partial}\phi_l. 
\end{equation}
Then we can compute the energy current as~\cite{Viti2015}
\begin{equation}
 \mathcal {J}_{LE}=\Tr[\rho_0 \Omega_0^{-1}\circ\Omega[T_l-\bar{T_l}] ],
\end{equation}
and observe that the cross-term in \eqref{thetabarT} vanishes because the initial state is factorized and the Hamiltonian satisfies the
symmetry $\partial \phi_{l/r}\rightarrow -\partial \phi_{l/r}$, and similarly for the antichiral component. We are left with
\begin{equation}
 \mathcal{J}_{LE}=\cos^2\gamma(\Tr[\rho_l \bar{T}_l]-\Tr[\rho_r T_r])=\frac{\pi\cos^2\gamma}{12}(\beta_l^{-2}-\beta_r^{-2}),
\end{equation}
where the last equality follows by standard CFT mapping on a cylinder.

%%%%%%%%%%%%%%%%%%%%%%%%%%%%%%%%%%%%%%%%%%%%
%
%
%%%%%%%%%%%%%%%%%%%%%%%%%%%%%%%%%%%%%%%%

\section{Finite-temperature matrix product states}
\label{ssec:method}

The transport protocol under study requires to prepare two (ideally) semi-infinite spin chains at given distinct temperatures $T_{l,r} = \beta^{-1}_{l,r}$ (we use units of $\hbar = k_B = 1$).
After the two halves have been connected, we compute the time evolution of the whole system.
We face this problem using an algorithm which exploits the so-called {\it ancilla method} in the context of matrix product states (MPS)~\cite{white2005,schollwock}. 
This method allows to represent the mixed thermal states of a $d^N$ dimensional Hilbert space (system) 
as pure states (written as MPS) on a $d^{2N}$ Hilbert space (system + ancilla). 

%%%%%%%%%%%%%%%%%%%%%%%%%%%%%%%%%%%%%%%%
%
\begin{figure}[t]
 \includegraphics[width=\columnwidth]{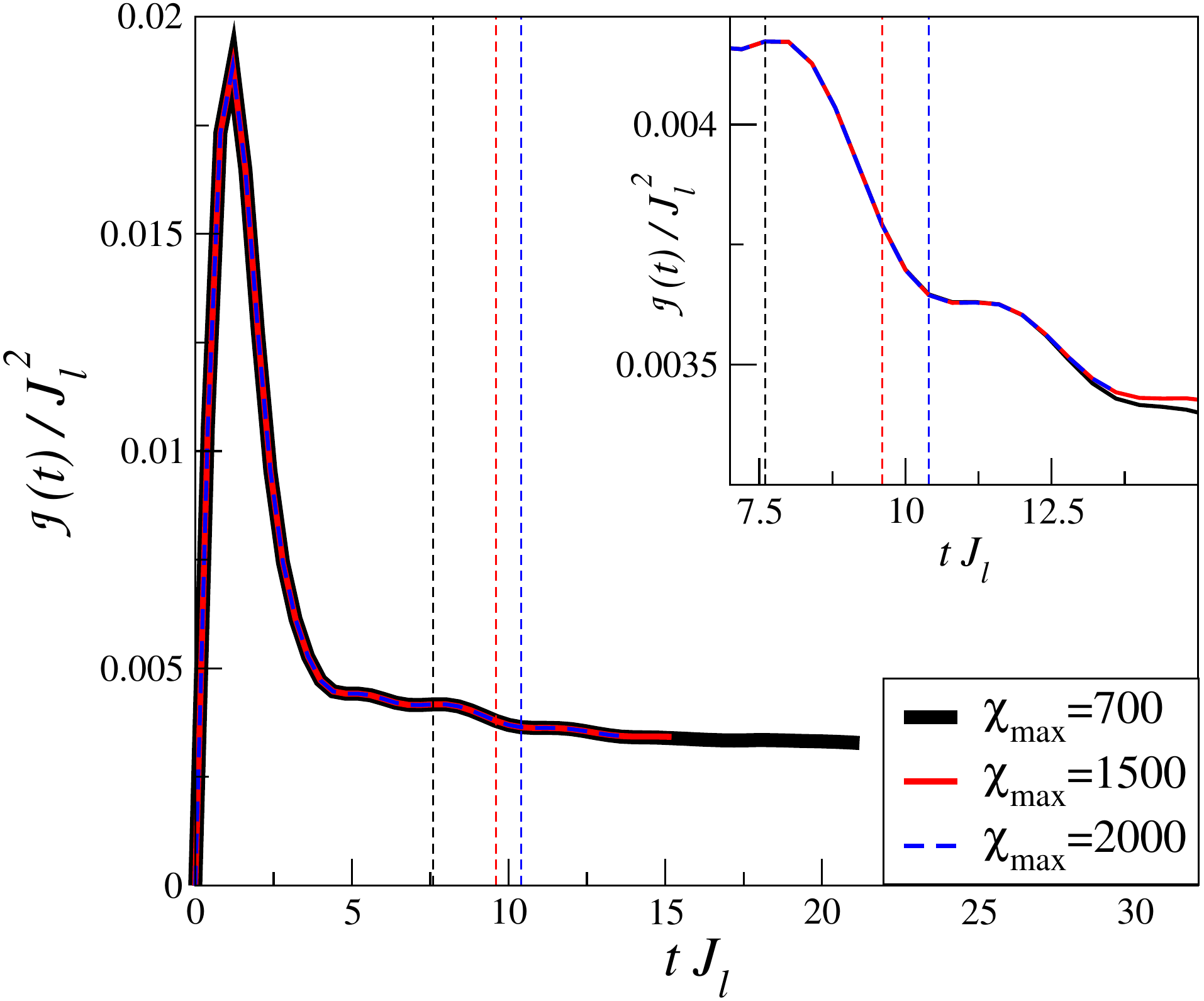}
 \caption{(color online) Time evolution of the energy current $\mathcal{J}(t)$ for different values of  $\chi_{\rm max}$, as indicated in the legend.
The dashed vertical lines are the times $t_{\rm ex}(\chi_{\rm max})$ after which we approximate the time-evolution according to the value of $\chi_{\rm max}$.
Here the parameters are $\beta_l=4 \ J_l^{-1}$, $\beta_r=5 \ J_l^{-1}$, $\Delta_l=-\Delta_r=-0.3$ and $J_r/J_l=1$.
In the inset we show a magnification of the data around the times $t_{\rm ex}(\chi_{\rm max})$.}
\label{fig:chi_approx}
\end{figure}
%
%%%%%%%%%%%%%%%%%%%%%%%%%%%%%%%%%%%%%%%% 

Let us start from the fact that any thermal density matrix $\hat\rho(\beta)={\rm e}^{-\beta\hat H_{\text{th}}}/\mathcal Z(\beta)$ 
($\mathcal Z(\beta)= {\rm Tr}[ {\rm e}^{-\beta \hat H_{\text{th}}}]$ being the partition function) can be written as
\begin{equation}
  \hat\rho(\beta) = \frac{\mathcal Z(0)}{\mathcal Z(\beta)} \ {\rm e}^{-\beta\hat H_{\text{th}}/2} \ \hat\rho(0) \ {\rm e}^{-\beta\hat H_{\text{th}}/2}, 
\end{equation}
where $\hat\rho(0)=\hat {\mathbb{I}}/\mathcal Z(0)$ is the (fully mixed) state of the system at infinite temperature, $\beta=0$. 
The state $\hat\rho(0)$ can be obtained considering a pure maximally entangled state \mbox{$\ket{\psi_{\rm ME}}=\bigotimes_{i=1}^N\left( \sum_{j=1}^d \frac1{\sqrt{d}} \ket{j^{(i)}}_S\ket{j^{(i)}}_A \right) $} in an enlarged Hilbert space made by the system ($S$) and a copy of it (the ancilla $A$) after one traces out the ancillary degrees of freedom.
Here $\big\lbrace \ket{j^{(i)}}_{S,A} \big\rbrace$ are the local basis elements of the $d$-dimensional local Hilbert space of the system and of the ancilla.
In this way we get 
\begin{equation}
  \label{ancilla02}
  \hat\rho(\beta) = \frac{\mathcal Z(0)}{\mathcal Z(\beta)} {\rm Tr}\left[  \ket{\psi_\beta} \! \bra{\psi_\beta} \right], \quad \mbox{with } \ket{\psi_\beta}\!=\!{\rm e}^{-\frac{\beta}{2}\hat H_{\text{th}}}\! \ket{\psi_{\rm ME}}.
\end{equation}

In terms of MPS, the maximally entangled state $\ket{\psi_{\rm ME}}$ is readily generated variationally, since it is a MPS (in the composite Hilbert space $S+A$) with bond-link dimension $\chi=1$.
The state $\ket{\psi_\beta}$ is then obtained by means of an imaginary-time evolution.
As $\beta$ increases, $\ket{\psi_\beta}$ gets entangled and the bond-link dimension needed for its representation increases.
In order to simulate $\hat\rho_0$ in Eq.~\eqref{initial_state}, one has to set $\hat H_{\text{th}} = \hat H_l + (\beta_r/\beta_l) \ \hat H_r$ and $\beta=\beta_l$.
Once the initial thermal state is prepared, we let the system evolve under the action of $\hat H$ in Eq.~\eqref{eq:Hamiltonian} by exploiting a standard time-evolving block-decimation (TEBD) algorithm~\cite{Vidal2004, Daley_2004, White_2004, schollwock}.
In principle one has the freedom to choose the Hamiltonian $\hat H_A$ under which the ancilla evolves, but it turns out that the careful choice $\hat H_A=-\hat H$ is fundamental to reach sufficiently long times~\cite{kennes2014,moore2012}.

In all the data that we show, the bond link is $\chi\le1000$ and the system size is such that, for the times explored, the dynamics does is not affected by the boundaries (we always simulate finite systems with open boundary conditions).

As mentioned in the main text, the bond link $\chi$ needed to faithfully represent the state of the system increases exponentially in time because of the spreading of the correlations.
Since a larger bond link means a larger computational time to evolve of a time-step,
in order to reach sufficiently long times we have to fix an upper bound $\chi_{\rm max}$ to the bond link value.
This means that, after a certain time $t_{\rm ex}(\chi_{\rm max})$, we start to approximate the {\it real} state of the system by {\it truncating} the bond link dimension to $\chi_{\rm max}$ (note that the TEBD procedure already introduces an error in the representation of the time-evolved wavefunction).

The operative way that we use in order to keep such approximation under control is to check the expectation value of the observables in which we are interested.
For example, given $\chi_{\rm max}$, if for $t>t_{\rm ex}(\chi_{\rm max})$  
the expectation value of the energy current operator~\eqref{eq:current0} 
(approximately) does not change as the bond link dimension is increased beyond $\chi_{\rm max}$, we can conclude that the correlations we are neglecting with our approximate representation are not relevant to describe the observable.

This heuristic method is exemplified in Fig.~\ref{fig:chi_approx}.
The energy current as a function of time (for typical values of the parameters) is plotted for different values of $\chi_{\rm max}$, as indicated in the legend.
At a first glance, we note that the data in the main panel almost overlap. This is confirmed by the magnification shown in the inset.
To be more quantitative, at $t =13.6 J_l^{-1}$ (the common largest time explored), the relative difference between the data with $\chi_{\rm max}=1500$ and $\chi_{\rm max}=2000$ is about $0.008\%$, while it is $0.39\%$ between the data with $\chi_{\rm max}=700$ and $\chi_{\rm max}=2000$.

\end{document}